\documentclass[prl, twocolumn, superscriptaddress, showpacs]{revtex4-1}
\usepackage[plainpages=false,pdfpagelabels,colorlinks=true,linkcolor=blue,urlcolor=blue,citecolor=blue,pdftitle={},pdfauthor={},pdfdisplaydoctitle=true,pdfduplex=DuplexFlipLongEdge]{hyperref}
\usepackage{url}
\usepackage{amsmath}
\usepackage{amssymb,amsfonts}
\usepackage{graphicx}
\usepackage[utf8]{inputenc}
\usepackage{multirow}
\usepackage{color}
\usepackage[toc,page]{appendix}
\usepackage{color}

\usepackage[all]{xy}

\usepackage[normalem]{ulem}
\renewcommand\sout{\bgroup\markoverwith{\textcolor{red}{\rule[0.5ex]{4pt}{0.8pt}}}\ULon}

\newcommand{\ii}{\mathrm{i}}

\newcommand{\ket}[1]{| #1 \rangle}
\newcommand{\bra}[1]{\langle #1 |}
\newcommand{\be}{\begin{equation}}
\newcommand{\ee}{\end{equation}}
\newcommand{\Su}{\mathrm{SU}(2)}

\newcommand{\rom}[1]{\uppercase\expandafter{\romannumeral #1\relax}}

\newcommand{\Tr}{\mathrm{Tr}}

\newcommand{\bse}{\begin{subequations}}
\newcommand{\ese}{\end{subequations}}

\newcommand{\bracket}[2]{\langle{#1}\vert{#2}\rangle}

\newcommand{\bpm}{\begin{pmatrix}}
\newcommand{\epm}{\end{pmatrix}}
\newcommand{\bmm}{\begin{matrix}}
\newcommand{\emm}{\end{matrix}}

\newcommand{\x}{\times}

\allowdisplaybreaks

\makeatletter
\newcommand*{\Relbarfill@}{\arrowfill@\Relbar\Relbar\Relbar}
\newcommand*{\xeq}[2][]{\ext@arrow 0055\Relbarfill@{#1}{#2}}
\makeatother

\usepackage[normalem]{ulem}
\renewcommand\sout{\bgroup\markoverwith{\textcolor{red}{\rule[0.5ex]{4pt}{0.8pt}}}\ULon}

\def\be{\begin{eqnarray}}
\def\ee{\end{eqnarray}}

\def\mb{\left(\begin{matrix}}
\def\me{\end{matrix}\right)}

\newcommand{\Psix}[3][1]{
\begin{tikzpicture}[scale=0.8]
\node[name=s, regular polygon, regular polygon sides=6, minimum size=1cm, outer sep=0pt ,draw] at (0,0) {}; 
\foreach \anchor/\x/\y /\xx/\yy /\b in
{corner 1/0.17/0.17*1.732/-0.11/0.18/1, corner 2/-0.17/0.17*1.732/0.07/0.18/2, corner 3/-0.34/0/-0.15/-0.18/3, corner 4/-0.17/-0.17*1.732/-0.22/-0.05/4, corner 5/0.17/-0.17*1.732/0.2/-0.05/5, corner 6/0.34/0/0.15/-0.18/6}
{
 \draw[shift=(s.\anchor)] (0,0) -- (\x,\y) node at(\xx,\yy) {$#2_{\text{\scalebox{0.7}{$\b$}}}$};
 \ifnum #1=1
 \draw[shift=(s.\anchor),<-,>=stealth', line width=0.01pt] (s.\anchor) -- (\x,\y);
 \fi
 }
\foreach \anchor/\xx/\yy /\a in
{side 1/0/-0.18/1, side 2/-0.18/0.05/2, side 3/0.15/0.05/3, side 4/0/-0.18/4, side 5/-0.18/0.05/5, side 6/0.15/0.05/6}
 \draw[shift=(s.\anchor)]  node at(\xx,\yy) {$#3_{\text{\scalebox{0.7}{$\a$}}}$};
\ifnum #1=1{
  \foreach \anchorr/\anchorf in
   {corner 1/corner 2, corner 2/corner 3, corner 3/corner 4, corner 4/corner 5, corner 5/corner 6, corner 6/corner 1}
   \draw[shift=(s.\anchorr), ->, >=stealth', line width=0.01pt]  (s.\anchorr) -- (s.\anchorf);}
 \else {
  \foreach \anchorb/\anchorw in
   {corner 1/corner 2, corner 3/corner 4, corner 5/corner 6} {
   \node[fill=black, circle, minimum size=0, inner sep=0, outer sep=0, draw] at(s.\anchorb) {};
   \node[fill=white, circle, minimum size=0, inner sep=0, outer sep=0, draw] at(s.\anchorw) {};}
}
\fi
\end{tikzpicture}
}

\usepackage{varioref}

\begin{document}

\title{Observation of Two-Vertex Four-Dimensional Spin Foam Amplitudes with a 10-qubit Superconducting Quantum Processor}

\author{Pengfei Zhang}
\thanks{P. Z., Z. H., and C. S. contributed equally to this work.}
\affiliation{\mbox{Interdisciplinary Center for Quantum Information, State Key Laboratory of Modern Optical Instrumentation,} and Zhejiang Province Key Laboratory of Quantum Technology and Device, Department of Physics, Zhejiang University, Hangzhou 310027, China}
\author{Zichang Huang}
\thanks{P. Z., Z. H., and C. S. contributed equally to this work.}
\affiliation{State Key Laboratory of Surface Physics, Fudan University, Shanghai 200433, China}
\affiliation{Department of Physics, Center for Field Theory and Particle Physics, and Institute for Nanoelectronic devices and Quantum computing, Fudan University, Shanghai 200433, China}
\author{Chao Song}
\thanks{P. Z., Z. H., and C. S. contributed equally to this work.}
\affiliation{\mbox{Interdisciplinary Center for Quantum Information, State Key Laboratory of Modern Optical Instrumentation,} and Zhejiang Province Key Laboratory of Quantum Technology and Device, Department of Physics, Zhejiang University, Hangzhou 310027, China}
\author{Qiujiang Guo}
\affiliation{\mbox{Interdisciplinary Center for Quantum Information, State Key Laboratory of Modern Optical Instrumentation,} and Zhejiang Province Key Laboratory of Quantum Technology and Device, Department of Physics, Zhejiang University, Hangzhou 310027, China}
\author{Zixuan Song}
\affiliation{\mbox{Interdisciplinary Center for Quantum Information, State Key Laboratory of Modern Optical Instrumentation,} and Zhejiang Province Key Laboratory of Quantum Technology and Device, Department of Physics, Zhejiang University, Hangzhou 310027, China}
\author{Hang Dong}
\affiliation{\mbox{Interdisciplinary Center for Quantum Information, State Key Laboratory of Modern Optical Instrumentation,} and Zhejiang Province Key Laboratory of Quantum Technology and Device, Department of Physics, Zhejiang University, Hangzhou 310027, China}
\author{Zhen Wang}
\affiliation{\mbox{Interdisciplinary Center for Quantum Information, State Key Laboratory of Modern Optical Instrumentation,} and Zhejiang Province Key Laboratory of Quantum Technology and Device, Department of Physics, Zhejiang University, Hangzhou 310027, China}
\author{Hekang Li}
\affiliation{\mbox{Interdisciplinary Center for Quantum Information, State Key Laboratory of Modern Optical Instrumentation,} and Zhejiang Province Key Laboratory of Quantum Technology and Device, Department of Physics, Zhejiang University, Hangzhou 310027, China}
\author{Muxin Han}
\email{hanm@fau.edu}
\affiliation{Department of Physics, Florida Atlantic University, 777 Glades Road, Boca Raton, FL 33431, USA}
\affiliation{Institut f\"ur Quantengravitation, Universit\"at Erlangen-N\"urnberg, Staudtstr. 7/B2, 91058 Erlangen, Germany}
\author{Haohua Wang}
\email{hhwang@zju.edu.cn}
\affiliation{\mbox{Interdisciplinary Center for Quantum Information, State Key Laboratory of Modern Optical Instrumentation,} and Zhejiang Province Key Laboratory of Quantum Technology and Device, Department of Physics, Zhejiang University, Hangzhou 310027, China}
\author{Yidun Wan}
\email{ydwan@fudan.edu.cn}
\affiliation{State Key Laboratory of Surface Physics, Fudan University, Shanghai 200433, China}
\affiliation{Department of Physics, Center for Field Theory and Particle Physics, and Institute for Nanoelectronic devices and Quantum computing, Fudan University, Shanghai 200433, China}
\affiliation{Department of Physics and Institute for Quantum Science and Engineering, Southern University of Science and Technology, Shenzhen 518055, China}

\begin{abstract}
Quantum computers are an increasingly hopeful means for understanding large quantum many-body systems bearing high computational complexity. Such systems exhibit complex evolutions of quantum states, and are prevailing in fundamental physics, e.g., quantum gravity. Computing the transition amplitudes between different quantum states by quantum computers is one of the promising ways to solve such computational complexity problems. In this work, we apply a 10-qubit superconducting quantum processor, where the all-to-all circuit connectivity enables a many-body entangling gate that is highly efficient for state generation, to studying the transition amplitudes in loop quantum gravity. With the device metrics such as qubit coherence, control accuracy, and integration level being continuously improved, superconducting quantum processors are expected to outperform their classical counterparts in handling many-body dynamics and may lead to a deeper understanding of quantum gravity.
\end{abstract}

\date{\today}
\maketitle

A pivotal question that any quantum theory should address is to predict the evolution of quantum states. In Spin Foam Model (SFM)~\cite{Reisenberger:1996pu,rovelli2014covariant,Perez2012,Barrett:2009mw,CFsemiclassical,HZ,HZ1,Han:2017xwo,Han:2018fmu,Han:2016fgh}---a covariant formulation of Loop Quantum Gravity (LQG)~\cite{book,Han:2005km,review1}, a 3-dimensional space is described by a quantum state---3-dimensional space state. In SFM, the pivotal question is answered by spin foam amplitudes, which are the probability amplitudes of 4-dimensional quantum spacetime regions formed by the evolution from initial 3-dimensional space states to final 3-dimensional space states. Calculating spin foam amplitudes is one of the crucial steps of applying the SFM to many interesting topics in LQG, e.g., the Planck star tunneling, blackhole-whitehole transition, cosmology, etc~\cite{Haggard_2015,Rovelli_2014,Christodoulou_2016,christodoulou2018characteristic,han2020semiclassical,ASHTEKAR2009347}. Unfortunately, computing spin foam amplitudes for general quantum spacetimes on a classical computer is numerically difficult and resource consuming~\cite{Don__2018,Don__2019,dona2020numerical}. Nevertheless, such numerical difficulty may be circumvented if we compute a spin foam amplitude on a quantum computer, which obtains the amplitude by experimental measurements~\cite{Li_2019,mielczarek2018spin,czelusta2020quantum}. 

Superconducting circuits provide a competitive solution for building a practical quantum computer. Superconducting qubits are patterned lithographically and controlled by precisely-assembled microwave pulses, with a rich parameter space of qubit properties and operation regimes experimentally accessible. Thanks to recent improvements in qubit coherence~\cite{Gustavsson_2016, Pop_2014, Chang_2013, Bruno_2015}, which limits the participation in energy storage for the circuit regions with two-level state defects~\cite{Wang_2009}, superconducting qubits have gained rapid developments in both science and engineering, covering various aspects including decoherence mechanisms~\cite{Pop_2014, Andras_2019}, quantum gates and algorithms~\cite{Chow_2011, Guo_2018, Barkoutsos_2018}, entanglement manipulations~\cite{Omran_2019, Song_2019, Barends_2014, DiCarlo_2010}, error-correction codes~\cite{Kandala_2019, Reed_2012, Andersen_2014}, and quantum simulations~\cite{Roushan_2017, Kandala_2017, Xu_2018, Song_2018, Georgescu_2014}. More recently, a superconducting quantum processor with 53 qubits was used to implement random one- and two-qubit gates for quantum chaos~\cite{Arute_2010}, which signifies the entrance to the era of noisy intermediate-scale quantum technologies for quantum speedup.

In this paper, we employ a 10-qubit superconducting quantum processor 
to compute the Ooguri spin foam amplitudes~\cite{doi:10.1142/S0217732392004171} given various 3-dimensional space states. We generate a basic quantum spacetime state, a 5-qubit single-vertex state, with the state fidelity as high as $0.832\pm0.005$, which we then duplicate to generate two 5-qubit single-vertex states in parallel, and apply a two-qubit entangling gate to emulate the `gluing' operation of the two single-vertex states to obtain a 10-qubit two-vertex state. We measure the spin foam amplitudes by taking the inner product of these vertex states and any given spin-$\frac{1}{2}$ 3-dimensional space state. The measured spin foam amplitudes of the transitions from a 3-dimensional space state consisting of $m$ regular quantum tetrahedra to a 3-dimensional space state consisting of $5-m$ quantum tetrahedra for the single-vertex state ($8-m$ quantum tetrahedra for the two-vertex state) agree with theory decently.
For various given $3$-dimensional space states, we find that in both the single-vertex and two-vertex cases, the largest spin foam amplitude is always achieved by the $3$-dimensional space states that corresponds to the $3$-dimensional boundaries of classical $4$-dimensional simplicial complexes, which are geometric objects widely used as discretized spacetime~\cite{Simplical,regge,Lawphongpanich2001}.

\begin{figure}[htbp]
        \centering\includegraphics[width=3.4in]{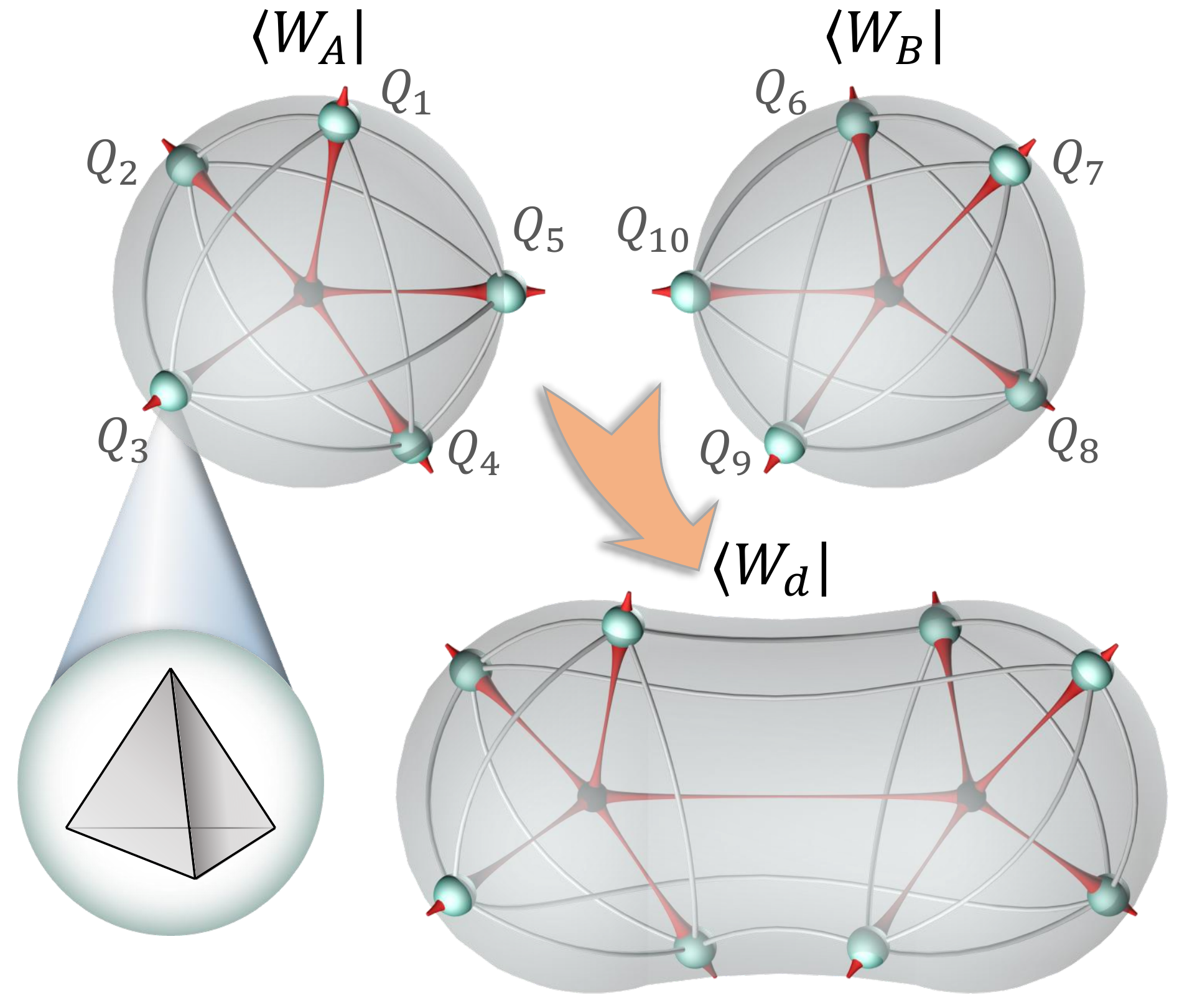}
        \caption{\textbf{Illustrative description of quantum spacetime} 
                Illustration of `gluing' spacetime atoms (top), assigned the states $\bra{W_{A}}$ and $\bra{W_{B}}$, for a bigger spacetime region as the two-vertex state $\bra{W_d}$ (bottom). Atom $A$ is bounded by five quantum tetrahedra $Q_1,\cdots, Q_5$, while $B$ is bounded by quantum tetrahedra $Q_6,\cdots, Q_{10}$. As shown in the lower left corner, each quantum tetrahedron is made of four spin states representing the tetrahedron's four quantized triangular faces. The `gluing' procedure entangles the common boundary tetrahedra $Q_5$ and $Q_{10}$. As a result, there are only eight boundary quantum tetrahedra for the  larger spacetime region $\bra{W_d}$.
        }\label{fig:qtetra}
\end{figure}

The spin foam amplitude of a $4$-dimensional quantum spacetime region can be considered as a combination of the spin foam amplitudes of the basic building blocks---spacetime atoms---of the quantum spacetime region. The $3$-dimensional boundary space state of a spacetime atom comprises five quantum tetrahedra. In SFM, each quantum tetrahedron is a quantum spin state corresponding to a closed tetrahedron (see Appendices Section~{2.1}). In the case of a spin-$\frac{1}{2}$ quantum tetrahedron, it is described by a Bloch-sphere state of a qubit (see Appendices Section~{2.2}). A spacetime atom can be regarded as a quantum  process~\cite{rovelli2014covariant} for $m$ quantum tetrahedra evolving to $5-m$ quantum tetrahedra. The detailed calculation (Appendices Section~{3}) shows that the spin foam amplitude of the process only linearly depends 
on the tensor product, denoted by $\ket{\Phi}$, of  the five boundary quantum tetrahedra.
Hence, one can universally define a quantum state, i.e., a vertex state, such that the spin foam amplitude of a spacetime atom is given by the inner product $\langle W | \Phi \rangle$ between the vertex state, associated with $\ket{W}$, and the atom's boundary tensor state. 
A pair of spacetime atoms can be `glued' into a bigger region of the spacetime, described by a two-vertex state $\ket{W_d}$, via entangling the common boundary quantum tetrahedra of the two spacetime atoms, as depicted in Fig.~\ref{fig:qtetra}. 
Therefore, the spin foam amplitude of the spacetime region is given by the inner-product between the two-vertex state and the region's 3-dimensional boundary tensor state. 

In general, a quantum spacetime region is made by gluing many spacetime atoms, which form a highly entangled many-body state~\cite{oriti2017spacetime,Hu_2005,Kotecha_2018,oriti2018bronstein,Chirco_2019,chirco2019generalized,Kotecha_2019,assanioussi2020thermal,Dong_2018,Pastawski_2015,Han:2016xmb,Han:2018fmu,chirco2019group,Chirco_2018,Asante_2019,Livine_2018}. By generating such a state, a quantum processor is capable of calculating the spin foam amplitude of any given boundary state up to a phase factor.

Here we utilize a superconducting quantum processor for a proof-of-concept demonstration with the boundary states made of spin-$\frac{1}{2}$ quantum tetrahedra, for which the basic spacetime atom is described by a 5-qubit state with its components given in Tab.~\ref{Stab_vertex_state_coefficients}. Our experiment bridges the fundamental concepts in LQG and the state-of-the-art technology of superconducting qubits, with consistent and enlightening results, and points out a promising path towards computing spin foam amplitude for a spacetime containing arbitrary number of spacetime atoms bounded by quantum tetrahedra with arbitrary spins , which could be useful for the topics aforementioned, e.g., the Planck-star tunneling, black hole-white hole transition, etc.

\begin{table}[!htbp]
        \scriptsize
        \addtolength{\tabcolsep}{+0.8pt}
        \renewcommand\arraystretch{2}
        \begin{tabular*}{3.4in}{c c c c c c c c}
                \hline
                \hline
                $\ket{00000}$ & $\ket{00001}$ & $\ket{00010}$ & $\ket{00011}$ & $\ket{00100}$ & $\ket{00101}$ & $\ket{00110}$ & $\ket{00111}$ \\
                $3\sqrt{3}$   & $9$           &           $9$ & $-3\sqrt{3}$ &
                $9$           & $9\sqrt{3}$   &  $-3\sqrt{3}$ &          $3$ \\
                \hline
                $\ket{01000}$ & $\ket{01001}$ & $\ket{01010}$ & $\ket{01011}$ & $\ket{01100}$ & $\ket{01101}$ & $\ket{01110}$ & $\ket{01111}$\\
                $9$           & $9\sqrt{3}$   &   $9\sqrt{3}$ &         $-9$ &
                $-3\sqrt{3}$  & $-9$          &           $3$ &  $-\sqrt{3}$ \\
                \hline
                $\ket{10000}$ & $\ket{10001}$ & $\ket{10010}$ & $\ket{10011}$ & $\ket{10100}$ & $\ket{10101}$ & $\ket{10110}$ & $\ket{10111}$ \\
                $9$           & $-3\sqrt{3}$  &   $9\sqrt{3}$ &          $3$ &
                $9\sqrt{3}$   & $-9$          &          $-9$ &  $-\sqrt{3}$ \\
                \hline
                $\ket{11000}$ & $\ket{11001}$ & $\ket{11010}$ & $\ket{11011}$ & $\ket{11100}$ & $\ket{11101}$ & $\ket{11110}$ & $\ket{11111}$\\
                $-3\sqrt{3}$  & $3$           &          $-9$ &  $-\sqrt{3}$ &
                $3$           & $-\sqrt{3}$   &   $-\sqrt{3}$ &         $21$ \\
                \hline
                \hline
        \end{tabular*}
        \caption{\label{Stab_vertex_state_coefficients} \footnotesize{
                        Wavefunction coefficients of the ideal 5-qubit vertex state in the computational basis. All displayed coefficients should be divided by $8\sqrt{42}$ for normalization. 
        }}
\end{table}

\begin{figure*}
\centering
\includegraphics[width=7in]{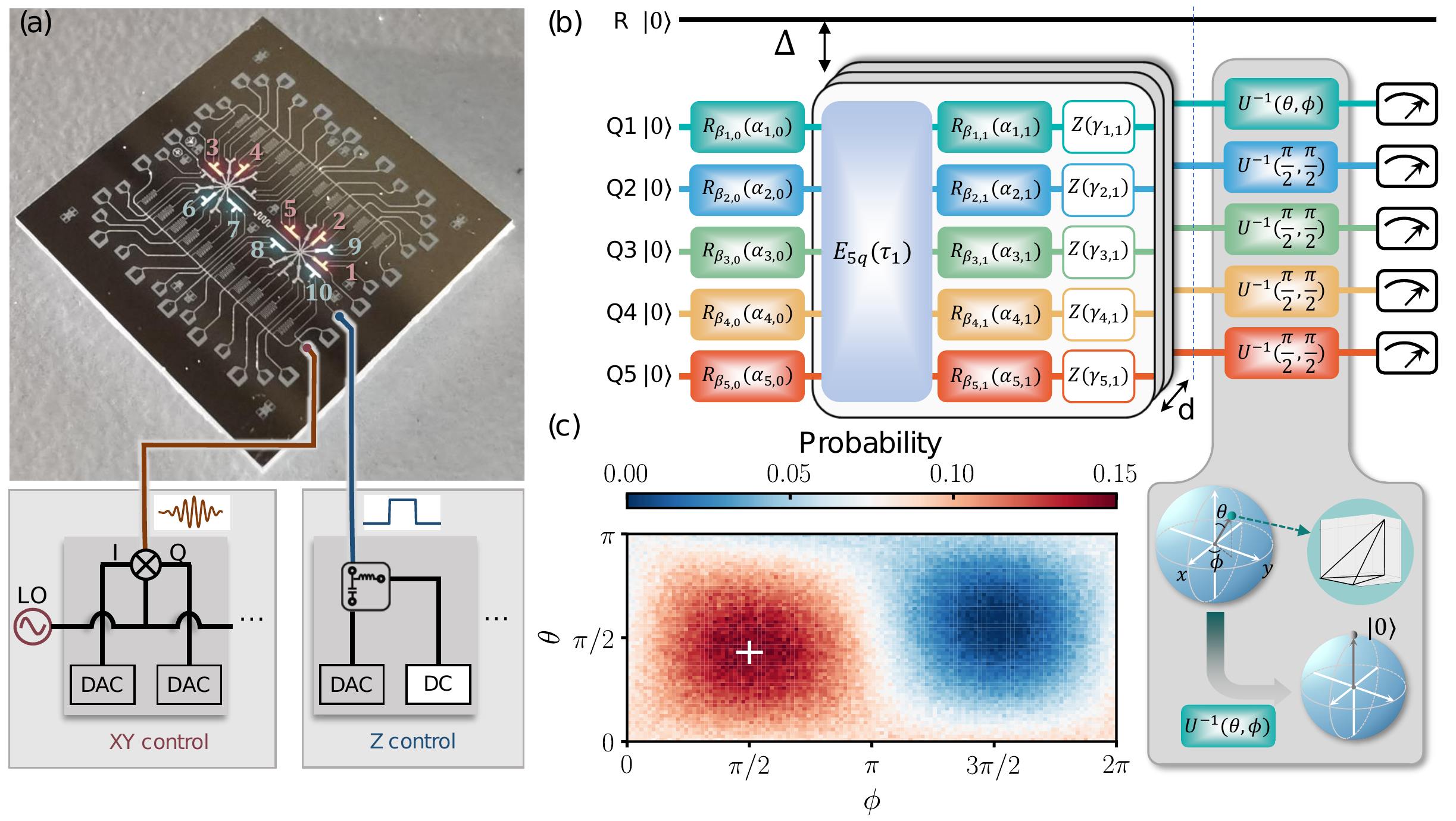}
\caption{\label{fig_single_vertex_state} \footnotesize{
\textbf{Device setup and the experiment of simulating a single spacetime atom.}
(a) Device image showing the 10 qubits highlighted and divided into two groups according to their coloring. Also illustrated are the external $XY$ control for implementing the single-qubit rotation $R_{{\beta}}(\alpha)$ by an angle $\alpha$ around ${\beta}$-axis, which is in the equator plane forming the angle $\beta$ with respect to $x$-axis, 
and the $Z$ control for realizing the phase gate $Z(\gamma)$, which rotates around $z$-axis by an angle $\gamma$.
(b) Sequence diagrams to generate the 5-qubit vertex state $\ket{W}$ (before the vertical dashed line) and 
measure the corresponding transition amplitude $\bracket{\Phi}{W}$, 
which consist of a layered execution of the many-body entangling gate 
$E_{5q}(\tau)$ and multiple single-qubit $XY$ rotations and $Z$ phase gates. Shaded box: The $U^{-1}(\theta,\phi)$ 
gate reverses a quantum tetrahedron state represented by the spherical coordinates ($\theta$, $\phi$) on the Bloch sphere to the north pole $\ket{0}$, which can be realized via an XY rotation $R_{\phi-\pi/2}(\theta)$. To estimate the experimental 5-qubit density matrix $\rho_\textrm{exp}$, tomographic operations (not shown) replacing the $U^{-1}$ gates followed by multiqubit readout are applied right after the generation of $\ket{W}$ (see Appendices Section~{1.4}). 
(c) Probability that all qubits are simultaneously measured to be in $\ket{0}$ resulting from the projection of $\ket{W}$ onto $\bra{\Phi}$, where four boundary tetrahedra of $\bra{\Phi}$ are fixed to be regular (i.e., $\theta=\pi/2$ and $\phi=\pi/2$) and the fifth one is continuously deformed as a function of $\theta$ and $\phi$. Maximum probability occurs at about $\theta=\pi/2$ and $\phi=\pi/2$, as identified at the crossing point. 
}}
\end{figure*}

Our superconducting quantum processor (Fig.~\ref{fig_single_vertex_state}(a)) is constructed by 20 frequency-tunable transmon qubits symmetrically coupled to a central bus resonator ($R$) fixed at $\omega_R/2\pi \approx 5.245$~GHz, where 10 qubits are selected for this experiment ($Q_{j}$ for $j=1$ to 10) and the rest are far detuned in frequency (also neglected for the experiment).
The 10-qubit circuit Hamiltonian is
\begin{align}
\frac{H_1}{\hbar} =& \omega_{R} a^{\dagger}a + 
\sum_{j=1}^{10} \left[ \omega_{j}(t) \ket{1_{j}} \bra{1_{j}} + g_{j} \left(\sigma_{j}^{+}a + \sigma_{j}^{-}a^{\dagger} \right) \right] \nonumber \\
&+ \sum_{i, j \in 10} g_{i, j} \left( \sigma_{i}^{+} \sigma_{j}^{-} + \sigma_{i}^{-} \sigma_{j}^{+} \right), \label{system_hamiltonian}
\end{align}
where $Q_j$'s resonance frequency $\omega_{j}(t)$ can be dynamically tuned within a couple of nanoseconds, $\sigma_{j}^{+}$ ($\sigma_{j}^{-}$) is the raising (lowering) operator of $Q_{j}$, 
$a^{\dagger}$ ($a$) the creation (annihilation) operator of $R$, $g_{j}$ the coupling strength between $Q_{j}$ and $R$, and $g_{i, j}$ direct coupling strength between $Q_{i}$ and $Q_{j}$. 
Idle frequencies of these 10 qubits are carefully arranged in order to minimize qubit-qubit crosstalks caused by simultaneous single-qubit rotation and readout pulses. Single-qubit $XY$ rotations, implemented by a 20-ns full-width at half maximum Gaussian-shape microwave pulse with a full length of 40~ns, as shown in Fig.~\ref{fig_single_vertex_state}(a) inset, are benchmarked to be above 0.990 in fidelity. $Z$ phase gates, implemented by fast square pulses, are used to dynamically tune qubit frequencies and to acquire desired dynamical phases, the latter of which is realized by appending additional phases to the axes of subsequent $XY$ rotations. Readout fidelity metrics are all above 0.95 (0.90) for the $\ket{0}$ ($\ket{1}$) states of these qubits, and are used to eliminate the measurement errors (Appendices Section~{1.1}). 

To implement the many-body entangling gate, the 10 qubits are divided into groups $A$ and $B$, and the qubits in either group are equally detuned from resonator $R$ by $\Delta_{A/B}=\omega_{A/B} - \omega_{R}$, where $\omega_{A} \neq \omega_{B}$, for a collective interaction mediated by $R$. The effective Hamiltonian in the dispersive regime with $R$ initially in the ground state is
\begin{align}
\frac{H_2}{\hbar} = \quad &\sum_{i, j \in A} \left(\frac{g_{i} g_{j} }{ \Delta_{A}} + g_{i, j} \right) \left( \sigma_{i}^{+} \sigma_{j}^{-} + \sigma_{i}^{-} \sigma_{j}^{+} \right) \nonumber\\
+ &\sum_{i, j \in B} \left(\frac{g_{i} g_{j} }{ \Delta_{B}} + g_{i, j} \right) \left( \sigma_{i}^{+} \sigma_{j}^{-} + \sigma_{i}^{-} \sigma_{j}^{+} \right) \nonumber\\
+ &\sum_{j \in A} \frac{g_{j}^2}{\Delta_{A}} \ket{1_{j}} \bra{1_{j}} + \sum_{j \in B} \frac{g_{j}^2}{\Delta_{B}} \ket{1_{j}} \bra{1_{j}}, \label{effective_system_hamiltonian}
\end{align}
where $g_{i} g_{j} / \Delta_{A/B} + g_{i, j}$ $(\equiv g_{i, j}^\textrm{Eff})$ is the effective intra-group coupling strength between $Q_i$ and $Q_j$. Therefore, the intra-group pairwise qubit couplings can be dynamically turned on by tuning the group of qubits 
on resonance with each other but detuned from $\omega_{R}$, which realizes the many-body entangling gate $E_{5q}(\tau)$ for the five qubits within the group, while the inter-group qubit-qubit couplings are effectively off provided that $\omega_{A}$ and $\omega_{B}$ are kept largely different.
        
Using $E_{5q}(\tau)$, we are able to deterministically generate the 5-qubit vertex state $\ket{W}$ by positioning the qubits in group $A$ at $\Delta_A/2\pi \approx -235$~MHz while biasing the qubits in group $B$ far detuned. 
As depicted in Fig.~\ref{fig_single_vertex_state}(b), the generation sequence is built upon the repeated execution of the many-body entangling gate with a variable duration $\tau_i$ followed by sequential single-qubit rotations $R_{\beta_{q,i}}\left(\alpha_{q,i}\right)$ and $Z(\gamma_{q,i})$, where $\alpha_{q,i}$ represents the rotation angle around $\beta_{q,i}$-axis, $\gamma_{q,i}$ denotes the rotation angle around $z$-axis, $q$ identifies the qubit, and $i$ ($= 1, \cdots, d$) refers to the depth position.
We first calibrate the effective intra-group coupling strengths $g_{i, j}^\textrm{Eff}$ 
using resonant qubit-qubit swap dynamics (Appendices Section~{1.2}), based on which we employ a gradient-based optimization procedure to acquire the values of $\tau_i$, $\alpha_{q,i}$, $\beta_{q,i}$, and $\gamma_{q,i}$,
in order to maximize the state fidelity of the target $\ket{W}$.
The layered execution of the many-body entangling gate is numerically shown to be highly effective in generating $\ket{W}$. Ideally, the parameter values for a state fidelity above 0.99 could be numerically acquired by implementing $d=4$ layers. Considering the finite coherence of our device, here we choose to prepare $\ket{W}$ with $d=2$ layers to shorten the sequence time:
We experimentally generate $\ket{W}$ and measure the 5-qubit density matrix $\rho_\textrm{exp}$ by quantum state tomography, yielding 
the state fidelity $\mathcal{F} = \Tr\left(\rho_\textrm{exp}\cdot\rho_\textrm{ideal}\right)=0.832 \pm 0.005$ (Appendices Section~{1.4}). 

Based on the generated $\ket{W}$, we can obtain the complex conjugate of the spin foam amplitude $\bracket{\Phi}{W}$ by quantum measurement (see Appendices Section~{4}), with $\bra{\Phi}$ denoting the boundary state made of five quantum tetrahedra. 
In our experiments, we fix four out of five boundary quantum tetrahedra to be regular and leave the fifth with arbitrary shape (see Appendices Section~{2.3}). The resulting probability is displayed in Fig.~\ref{fig_single_vertex_state}(c), showing that the maximum probability occurs when the boundary contains five regular tetrahedra with consistent orientation to form a 4-simplex (Appendices Section~{2.4}), and our measured probability agrees with theory well.

\begin{figure}
\centering
\includegraphics[width=3.4in]{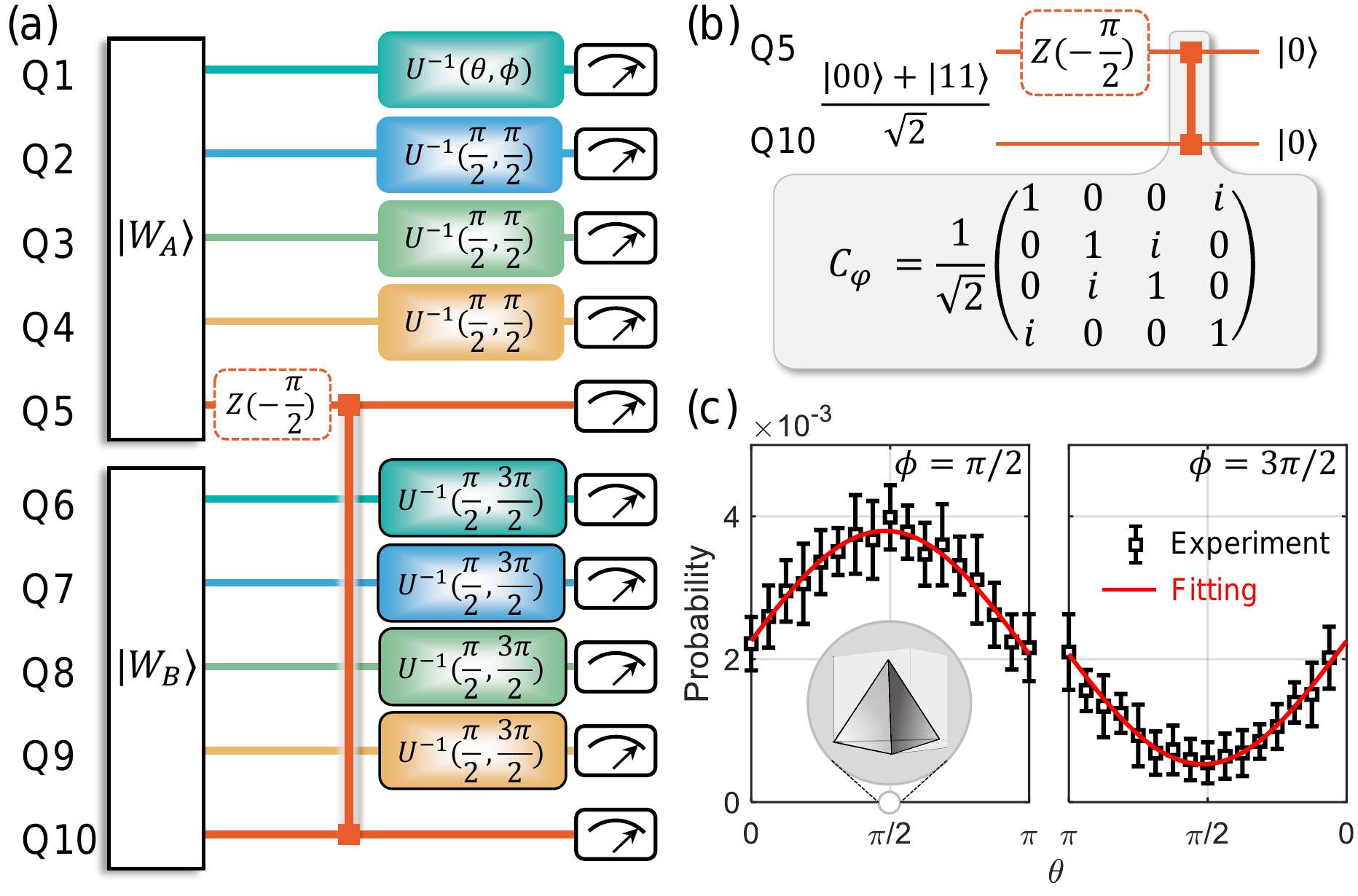}
\caption{\label{fig_double_vertex_state} \footnotesize{
\textbf{Experiment of simulating two spacetime atoms and their `gluing'.}
(a) Sequence diagrams to `glue' the two vertex states and measure the two-vertex four dimensional spin foam amplitude, where $\theta$ varies from 0 to $\pi$ with $\phi=\pi/2$ or $3\pi/2$. 
The $C_\varphi$ gate acting on $Q_5$-$Q_{10}$ transforms the  Einstein-Podolsky-Rosen (EPR) state $(\ket{00} + \ket{11})/\sqrt{2}$, 
representing the bulk tetrahedron in the two-atom boundary state, to $\ket{00}$. As in Fig.~\ref{fig_single_vertex_state}(b), 
the $U^{-1}(\theta,\phi)$ gates reverse the eight boundary tetrahedra qubits to $\ket{0}$. 
Seven out of the eight boundary tetrahedra are fixed as regular, and the qubits in groups $A$ and $B$ 
have reversed phases of $\phi =\pi/2$ and $3\pi/2$, for $\ket{W_{A}}$ and $\ket{W_{B}}$ to be `glued'.
(b) Illustration of transforming $(\ket{00} + \ket{11})/\sqrt{2}$ to $\ket{00}$ (Appendices Section~{1.5}).
(c) Probabilities of simultaneously measuring $\ket{0}$ for all qubits (squares with error bars) as a function of $\theta$, 
which parametrizes the shape of the eighth boundary tetrahedron. Lines are fits of the data to sinusoidal curves.
}}
\end{figure}

To generate two vertex-states in parallel involving all 10 qubits in both groups, we select $\Delta_A/2\pi \approx -295$~MHz and $\Delta_B/2\pi \approx -228$~MHz. Again here we employ a gradient-based optimization, with $d=2$, to acquire a sets of parameters for the 10 qubits, with the requirement  that the many-body entangling gate times $\tau_i$ are close for the two groups at the same depth position $i$. The qubits in the group with shorter $\tau_i$ are slightly detuned from each other after $\tau_i$, and subsequently synchronized with the qubits in the other group when moving to their corresponding idle frequencies for single-qubit operations (Appendices Section~{1.3}). These two simultaneously generated vertex states $\ket{W_{A}}$ and $\ket{W_{B}}$  have state fidelities $0.704 \pm 0.007$ and $0.722 \pm 0.015$.

We then `glue' $\ket{W_{A}}$ and $\ket{W_{B}}$ into a composite spacetime $\ket{W_d}$ via changing their common boundary tetrahedra into an EPR pair, $\left( \ket{00} + \ket{11}\right)/\sqrt{2}$, as illustrated in Figs.~\ref{fig_double_vertex_state}(a) and (b). This entanglement is
achieved by a two-qubit controlled phase ($C_\varphi$) gate executed 
in the dressed-state basis as demonstrated previously~\cite{PhysRevLett.121.130501}. The two-vertex state $\ket{W_d}$ characterizes the spin foams amplitude of a spacetime region as two spacetime atoms bounded by eight boundary quantum tetrahedra. Experimentally, we use $\ket{W_d}$ to calculate the spin foam amplitude of a set of boundaries where seven out of eight boundary quantum tetrahedra are regular and the eighth one arbitrary. Similar to the single-vertex case, the maximum probability occurs when all eight boundary quantum tetrahedra are regular and oriented consistently to form a boundary of two `glued' 4-simplices.

In conclusion, we have explored the SFM consisting of a single-vertex state and that of a two-vertex state on a superconducting quantum processor. We prepare the vertex states with a protocol relying on an efficient many-body entangling gate, taking the advantage of parallel operations generic with the multiqubit-resonator-bus architecture. 
Furthermore, we investigate the properties of quantum spacetime by collecting the varying spin foam amplitudes determined by the boundary quantum tetrahedra. This work not only actualizes the experimental viability on the study of quantum spacetime, that a quantum processor can automatically calculate the transition amplitudes of a spacetime made of multiple atoms, but also shows that superconducting quantum processors hold the promise of simulating and investigating complex dynamics of quantum many-body systems. 

\acknowledgements{
We acknowledge the support of the Natural Science Foundation of China (Grants No. 11725419 and No. 11875109), the National Key Research and Development Program of China (Grants No. 2017YFA0304300 and 2019YFA0308100), and the Zhejiang Province Key Research and Development Program (Grant No. 2020C01019).
Y.W. is also supported by Shanghai Municipal Science and Technology Major Project (Grant No. 2019SHZDZX01).}

\begin{appendices}

\section{1. Experimental details}
\subsection{1.1. Device}\label{sec:device}
The device used in this experiment is a 20-qubit superconducting quantum processor, 
where all qubits are capacitively coupled to the central bus resonator $R$. 
10 transmon qubits are selected for this experiment, labeled as $Q_{j}$ for $j=1$ to 10, with individual microwave control 
and flux bias lines for implementation of $XY$ and $Z$ rotations, respectively. 
Each qubit dispersively interacts with its own readout resonator, 
which is $\sim 1.5$~GHz above the qubit resonance and connected to one of the two transmission lines across the circuit chip. 
In practice, multi-tone microwave signals are passed through 
the transmission lines and scattered due to impedance mismatch as shunted by the multiple qubit-state-dependent readout resonators.  
The scattered signals are amplified and then demodulated at room temperature by analog to digital converters for detection of the multiqubit state.

Qubit idle frequencies $\omega_j$, where single-qubit operations are applied, are carefully arranged to minimize the crosstalk effect.
Single-qubit $X/2$ ($X$) and $Y/2$ ($Y$) gates are used to rotate qubit state on the Bloch sphere by $\pi/2$ ($\pi$) around $x$-axis and $y$-axis, respectively, 
and are characterized via simultaneous randomized benchmarking (RB), 
yielding gate fidelity values no less than 0.990 (Tabs.~\ref{Stab_qubit_characteristics_single_vertex_states} 
and \ref{Stab_qubit_characteristics_double_vertex_states}).
Also shown in these tables are the relevant qubit energy relaxation times at
the entangling frequency $\omega_{A/B}$ for the many-body entangling gate and those at $\omega_j$, as well as
the qubit readout fidelity metrics for the $\ket{0}$ ($\ket{1}$) states used to correct the directly
measured probabilities to eliminate the measurement errors~\cite{Zheng_2017}. 

\begin{table*}[!htbp]
	\addtolength{\tabcolsep}{+4pt}
	\renewcommand\arraystretch{1.5}
	\begin{tabular*}{\textwidth}{c | c c c c c c c c | c c}
		\hline
		\hline
		\quad    &   $\omega_{j}/2\pi$ &   Simultaneous &   Simultaneous &   Simultaneous &   Simultaneous &  $T_{1, j}$ & \multirow{2}{*}{$F_{0, j}$} & \multirow{2}{*}{$F_{1, j}$} & $\omega_{A}/2\pi$ & $T_{1, j}^\prime$\\
		\quad    &               (GHz) & $X/2$ fidelity &   $X$ fidelity & $Y/2$ fidelity &   $Y$ fidelity &    ($\mu$s) &                       \quad &                       \quad &             (GHz) &                   ($\mu$s)\\
		\hline
		$Q_{1}$  & 4.959 &  0.9974(2) &  0.9923(4) &  0.9962(2) &  0.9911(4) & 35.1 & 0.959 & 0.906 & \multirow{5}{*}{5.010} & 17.2 \\
		$Q_{2}$  & 5.001 &  0.9988(2) &  0.9937(4) &  0.9987(2) &  0.9926(4) & 28.3 & 0.969 & 0.928 &                  \quad & 47.8 \\
		$Q_{3}$  & 4.920 &  0.9986(1) &  0.9977(1) &  0.9988(1) &  0.9977(1) & 38.2 & 0.971 & 0.941 &                  \quad & 28.4 \\
		$Q_{4}$  & 5.053 &  0.9986(1) &  0.9957(2) &  0.9986(1) &  0.9953(2) & 31.9 & 0.983 & 0.934 &                  \quad & 39.5 \\
		$Q_{5}$  & 5.080 &  0.9975(1) &  0.9958(2) &  0.9976(2) &  0.9958(2) & 35.3 & 0.985 & 0.944 &                  \quad & 43.6 \\
		\hline
		\hline
	\end{tabular*}
	\caption{\label{Stab_qubit_characteristics_single_vertex_states} \footnotesize{
			Qubit characteristics for the generation of a single 5-qubit vertex state. 
			$\omega_{j}$ is the idle frequency of $Q_{j}$, where superexchange interactions are switched off and single-qubit rotations $R_{\beta}(\alpha)$ are implemented. 
			To switch on the intra-cluster superexchange interactions, qubits are biased to the corresponding interaction frequency $\omega_{A}$. 
			Each qubit's lifetime $T_{1,j}$ at both $\omega_{j}$ and corresponding $\omega_{A}$ are collected. 
			Performance of single-qubit $\pi/2$ and $\pi$ $XY$ rotations are verified by simultaneous randomized benchmarking. 
			To measure $T_{1,j}$, $Q_{j}$ is prepared to $\ket{1}$ by a $\pi$ $XY$ rotation, and then measurement is carried out after $t$. The corrected $\ket{1}$-state probability $P_{1, j}$ as a function of $t$ is fitted to $P_{1, j} \propto \mathrm{exp}(-t/T_{1, j})$. 
			While collecting each qubit's $T_{1, j}^\prime$ at $\omega_{A}$, other qubits are biased to around the sweet point. 
			The 5-qubit vertex state is denoted by $\ket{Q_{1}Q_{2}Q_{5}Q_{4}Q_{3}}$ with coefficients from Tab.~\ref{Stab_vertex_state_coefficients_1}. 
	}}
\end{table*}

\begin{table*}[!htbp]
	\addtolength{\tabcolsep}{+4pt}
	\renewcommand\arraystretch{1.5}
	\begin{tabular*}{\textwidth}{c | c c c c c c c c | c c}
		\hline
		\hline
		\quad    &   $\omega_{j}/2\pi$ &   Simultaneous &   Simultaneous &   Simultaneous &   Simultaneous &  $T_{1, j}$ & \multirow{2}{*}{$F_{0, j}$} & \multirow{2}{*}{$F_{1, j}$} & ${\omega_{A/B}}/{2\pi}$ & $T_{1, j}^\prime$ \\
		\quad    &               (GHz) & $X/2$ fidelity &   $X$ fidelity & $Y/2$ fidelity &   $Y$ fidelity &    ($\mu$s) &                       \quad &                       \quad &             (GHz) &                   ($\mu$s)\\
		\hline
		$Q_{1}$  & 4.580 &  0.9985(1) &  0.9986(1) &  0.9984(1) &  0.9989(1) & 31.7 & 0.962 & 0.928 & \multirow{5}{*}{4.950} & 25.4 \\
		$Q_{2}$  & 4.918 &  0.9979(1) &  0.9977(1) &  0.9981(1) &  0.9979(1) & 39.8 & 0.972 & 0.925 &                  \quad & 53.2 \\
		$Q_{3}$  & 4.731 &  0.9972(1) &  0.9951(1) &  0.9982(1) &  0.9980(1) & 41.5 & 0.957 & 0.901 &                  \quad & 29.6 \\
		$Q_{4}$  & 4.540 &  0.9989(1) &  0.9987(1) &  0.9988(1) &  0.9991(1) & 45.1 & 0.963 & 0.935 &                  \quad & 28.9 \\
		$Q_{5}$  & 5.080 &  0.9973(2) &  0.9910(4) &  0.9971(2) &  0.9913(4) & 29.8 & 0.985 & 0.925 &                  \quad & 37.0 \\
		\hline
		$Q_{6}$  & 4.646 &  0.9990(2) &  0.9978(2) &  0.9986(2) &  0.9978(1) & 28.3 & 0.970 & 0.907 & \multirow{5}{*}{5.017} & 34.1 \\
		$Q_{7}$  & 5.117 &  0.9983(1) &  0.9926(3) &  0.9981(1) &  0.9918(3) & 24.1 & 0.980 & 0.900 &                  \quad & 33.7 \\
		$Q_{8}$  & 4.960 &  0.9984(1) &  0.9966(2) &  0.9983(1) &  0.9965(1) & 50.8 & 0.973 & 0.918 &                  \quad & 49.5 \\
		$Q_{9}$  & 5.006 &  0.9972(2) &  0.9912(4) &  0.9979(3) &  0.9902(5) & 30.4 & 0.954 & 0.922 &                  \quad & 33.5 \\
		$Q_{10}$ & 5.043 &  0.9961(5) &  0.9985(3) &  0.9968(5) &  0.9979(3) & 40.1 & 0.960 & 0.918 &                  \quad & 38.2 \\
		\hline
		\hline
	\end{tabular*}
	\caption{\label{Stab_qubit_characteristics_double_vertex_states} \footnotesize{
			Qubit characteristics for the generation of two 5-qubit vertex states. Qubits in group $A$ and $B$ are biased to $\omega_{A}$ and $\omega_{B}$ to switch on the intra-cluster superexchange interactions respectively. More details about the symbols' meaning are provided previously in Tab.~\ref{Stab_qubit_characteristics_single_vertex_states}. 
			Here we slightly rearrange the qubits so that the two vertex states are given by $\ket{Q_{1}Q_{2}Q_{3}Q_{4}Q_{5}}$ and $\ket{Q_{6}Q_{7}Q_{8}Q_{9}Q_{10}}$ respectively. 
	}}
\end{table*}

\begin{figure}[!htbp]
	\centering
	\includegraphics[width=0.4\textwidth]{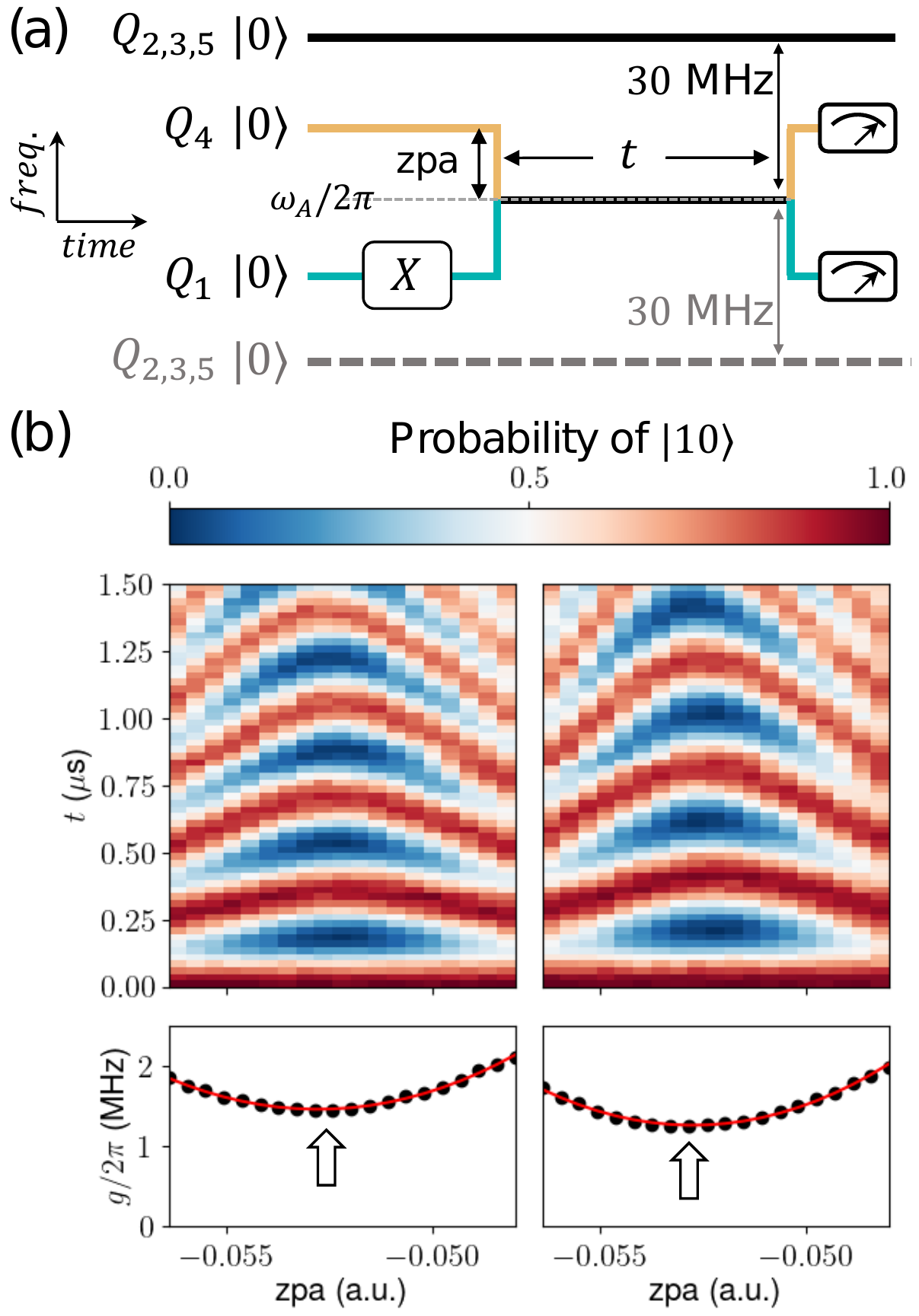}
	\caption{\label{Sfig_qqswap2D}\footnotesize{
			(a) Pulse sequences to measure the effective coupling strength between $Q_1$ and $Q_4$ at $\omega_A$, 
			while $Q_2$, $Q_3$, and $Q_5$ are positioned 30~MHz either above (black line) or below (gray line) $\omega_A/2\pi$. 
			The pulse amplitude (zpa) and the length ($t$) of the square $Z$ pulse applied to the flux bias line of $Q_{4}$ are varied during the measurement. 
			(b) The corrected probabilities of measuring $\ket{Q_1Q_4}$ in $\ket{10}$ with $Q_2$, $Q_3$, and $Q_5$ at 30~MHz above (left) and below (right) $\omega_A/2\pi$. 
			Arrows identify the minimum positions where $Q_1$ and $Q_4$ are on resonance. 
	}}
\end{figure}			

\subsection{1.2. Effective coupling strength $g_{i, j}^\textrm{Eff}$}
We use the two-qubit swap dynamics to estimate the effective coupling strength 
$g_{i, j}^\mathrm{Eff}$ between $Q_i$ and $Q_j$ within group $A$ 
at the entangling frequency $\omega_{A}$, while the 5 qubits in group $B$ are far detuned. 
As illustrated by the pulse sequence in Fig.~\ref{Sfig_qqswap2D}(a), 
we first excite $Q_{i}$ to $\ket{1}$ with an $X$ rotation, 
and then bias it to $\omega_{A}$ while $Q_{j}$ is swept 
across $\omega_{A}$ via a detuning pulse for a duration of $t$. 
The probability data of measuring $\ket{Q_iQ_j}$ in $\ket{10}$ as 
functions of the amplitude of the detuning pulse and $t$ 
resulting from the abovementioned swap dynamics during which the other three qubits 
are statically positioned 30~MHz above (left panel) and below (right panel) 
$\omega_A/2\pi$, are shown in Fig.~\ref{Sfig_qqswap2D}(b).
Both cases demonstrate well-resolved Chevron patterns.
For each column of the measured probabilities, we extract the oscillation frequency $2g$ by Fourier transform, 
and take the minimum $g$ value averaging over both cases as a close estimate of $g_{i, j}^\textrm{Eff}$. 
The 10 pairwise coupling strengths for the qubits in group $A$ used in the experiment of simulating 
a single spacetime atom in Fig.~2 of the main text are listed as values in brackets in Tab.~\ref{Stab_coupling_strength}. 

Similarly, to obtain the pairwise coupling strengths in the experiment of simulating two spacetime atoms, 
we measure the 10 pairwise coupling strengths for the qubits in one group 
following the abovementioned procedure while positioning the qubits in the other group at
their own entangling frequency. All intra-group pairwise coupling strengths are listed in Tab.~\ref{Stab_coupling_strength}. 
\begin{table*}[!htbp]
	\addtolength{\tabcolsep}{+6pt}
	\renewcommand\arraystretch{2}
	\begin{tabular*}{\textwidth}{c | c c c c c c c c c c}
		\hline
		\hline
		\quad & Q$_{1}$-Q$_{2}$ & Q$_{1}$-Q$_{3}$ & Q$_{1}$-Q$_{4}$ & Q$_{1}$-Q$_{5}$ & Q$_{2}$-Q$_{3}$ & Q$_{2}$-Q$_{4}$ & Q$_{2}$-Q$_{5}$ & Q$_{3}$-Q$_{4}$ & Q$_{3}$-Q$_{5}$ & Q$_{4}$-Q$_{5}$\\
		\hline
		\multirow{2}{*}{$g_{i, j}^{\mathrm{Eff}}/2\pi$ (MHz)} & -0.58 & -1.04 & -1.02 & -0.72 & -1.10 & -1.08 & -0.46 & -0.64 & -1.06 & -1.04\\
		\quad  &(-0.70)&(-1.33)&(-1.36)&(-0.99)&(-1.34)&(-1.36)&(-0.66)&(-0.81)&(-1.25)&(-1.26)\\
		\hline
		\hline
		\quad & Q$_{6}$-Q$_{7}$ & Q$_{6}$-Q$_{8}$ & Q$_{6}$-Q$_{9}$ & Q$_{6}$-Q$_{10}$ & Q$_{7}$-Q$_{8}$ & Q$_{7}$-Q$_{9}$ & Q$_{7}$-Q$_{10}$ & Q$_{8}$-Q$_{9}$ & Q$_{8}$-Q$_{10}$ & Q$_{9}$-Q$_{10}$\\
		\hline
		$g_{i, j}^{\mathrm{Eff}}/2\pi$ (MHz) & -0.74 & -1.22 & -1.39 & -1.39 & -1.15 & -1.30 & -1.30 & -1.00 & -1.01 & -0.67\\
		\hline
		\hline
	\end{tabular*}
	\caption{\label{Stab_coupling_strength} \footnotesize{
			Effective coupling strengths $g_{ij}^\textrm{Eff}$ estimated using the two-qubit swap dynamics. 
			Values in parentheses are for the single 5-qubit vertex state experiment. 
	}}
\end{table*}

\subsection{1.3. Generation of the 5-qubit vertex state}

\begin{table*}[!htbp]
	\scriptsize
	\addtolength{\tabcolsep}{+0.5pt}
	\renewcommand\arraystretch{2}
	\begin{tabular*}{\textwidth}{c | c c c c c c c c c c c c c c c c}
		\hline
		\hline
		Basis &
		$\ket{00000}$ & $\ket{00001}$ & $\ket{00010}$ & $\ket{00011}$ & $\ket{00100}$ & $\ket{00101}$ & $\ket{00110}$ & $\ket{00111}$ &
		$\ket{01000}$ & $\ket{01001}$ & $\ket{01010}$ & $\ket{01011}$ & $\ket{01100}$ & $\ket{01101}$ & $\ket{01110}$ & $\ket{01111}$\\
		\hline
		Coeff. &
		$3\sqrt{3}$   & $9$           &           $9$ & $-3\sqrt{3}$ &
		$9$           & $9\sqrt{3}$   &  $-3\sqrt{3}$ &          $3$ &
		$9$           & $9\sqrt{3}$   &   $9\sqrt{3}$ &         $-9$ &
		$-3\sqrt{3}$  & $-9$          &           $3$ &  $-\sqrt{3}$ \\
		\hline
		\hline
		Basis &
		$\ket{10000}$ & $\ket{10001}$ & $\ket{10010}$ & $\ket{10011}$ & $\ket{10100}$ & $\ket{10101}$ & $\ket{10110}$ & $\ket{10111}$ &
		$\ket{11000}$ & $\ket{11001}$ & $\ket{11010}$ & $\ket{11011}$ & $\ket{11100}$ & $\ket{11101}$ & $\ket{11110}$ & $\ket{11111}$\\
		\hline
		Coeff. &
		$9$           & $-3\sqrt{3}$  &   $9\sqrt{3}$ &          $3$ &
		$9\sqrt{3}$   & $-9$          &          $-9$ &  $-\sqrt{3}$ &
		$-3\sqrt{3}$  & $3$           &          $-9$ &  $-\sqrt{3}$ &
		$3$           & $-\sqrt{3}$   &   $-\sqrt{3}$ &         $21$ \\
		\hline
		\hline
	\end{tabular*}
	\caption{\label{Stab_vertex_state_coefficients_1} \footnotesize{
			Wavefunction coefficients of the ideal 5-qubit vertex state in the computational basis. All displayed coefficients should be divided by $8\sqrt{42}$ for normalization. 
	}}
\end{table*}

The wavefunction coefficients of the ideal 5-qubit vertex state in the computational basis are displayed in Tab.~\ref{Stab_vertex_state_coefficients_1} (or Tab.~\rom{1} of the main text).
Despite the fact that an arbitrary multi-qubit quantum state can be prepared 
with a series of single-qubit gates and a single type of two-qubit gates such as the CNOT gates, 
it remains an active and challenging subject to reduce the quantum circuit complexity because of the presence of quantum control noises. 
For example, to generate a 5-qubit state described by $2^{5} - 1$ complex parameters,  
a known decomposition method requires 57 single-qubit rotations and 26 CNOT gates with a depth of 22,
which is quite resource-consuming considering the limited coherence performance of our device \cite{Plesch_2011}. 
In this experiment, we take an alternative approach in state preparation based on a many-body entangling gate $E_{5q}(\tau)$ 
realized by evolving the on-resonant 5-qubit system for a variable duration $\tau$ 
under the Hamiltonian $H_{5q}$, where
\begin{equation*}
\label{coupling_hamiltonian}
H_{5q}/\hbar = \sum_{i < j} g_{i, j}^{\mathrm{Eff}} \left( \sigma_{i}^{+} \sigma_{j}^{-} + \sigma_{i}^{-} \sigma_{j}^{+} \right).
\end{equation*}
In principle, repetitive execution of this many-body entangling gate $E_{5q}(\tau)$ interlaced with arbitrary single-qubit gates can 
efficiently produce highly-entangled states such as the 5-qubit vertex state.

The sequence diagram to generate the 5-qubit vertex state is illustrated in Fig.~2(b) of the main text,
which consists of an initialization stage with 5 XY rotations and a $d$-layer repetition stage, where each layer 
contains a many-body entangling gate $E_{5q}(\tau)$, 5 XY rotations, and 5 Z phase gates. 
Using the measured pairwise couplings $g_{i, j}^\textrm{Eff}$ in Tab.~\ref{Stab_coupling_strength}, 
we numerically search within a constrained parameter space to locate a local maximum in the state fidelity of the generated 5-qubit vertex state, 
sampling over 10 independent trials with random initial guesses for the highest fidelity achievable. 
The numerically obtained parameter values are then implemented to guide the experiment of generating the 5-qubit vertex state. 

\begin{figure*}[!htbp]
	\centering
	\includegraphics[width=\textwidth]{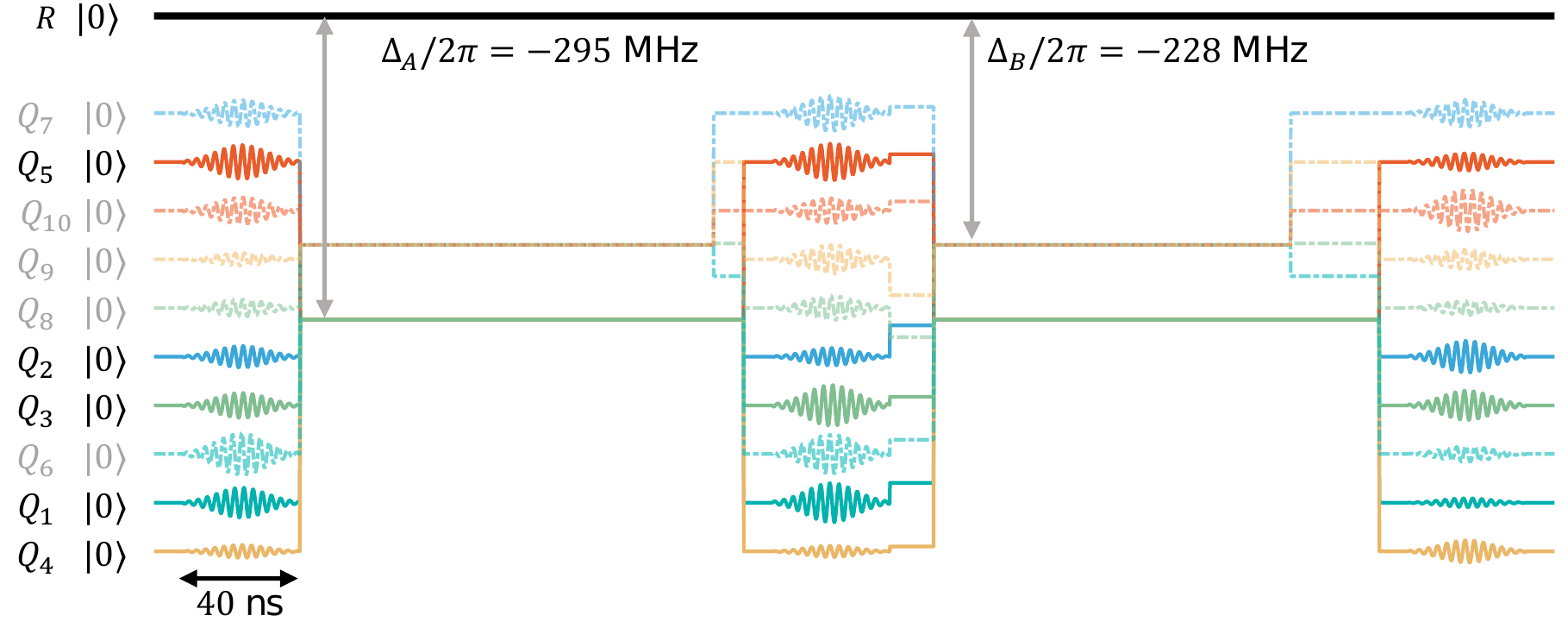}
	\caption{\label{Sfig_two_vertex_state_circuit} \footnotesize{
			Pulse sequence of generating two 5-qubit vertex states in parallel. 
			Qubits are separated into two groups $A$ and $B$ as differentiated by black and gray at the beginning, respectively. 
	}}
\end{figure*}

In the multiqubit-resonator-bus architecture, we are able to generate two 5-qubit vertex states $\ket{W_{A}}$ and $\ket{W_{B}}$ in parallel
by positioning two qubit groups $A$ and $B$ at different entangling frequencies, so that the inter-group qubit couplings are effectively turned off. 
The experimentally chosen entangling frequencies $\Delta_A$ and $\Delta_B$ differ by about $2\pi~\times~67$~MHz (see Tab.~\ref{Stab_qubit_characteristics_double_vertex_states}),
and a constraint that requires the many-body entangling durations of the two groups to be close is enforced in the numerical optimization.
The exact pulse sequence of generating $\ket{W_{A}}$ and $\ket{W_{B}}$ is shown in Fig.~\ref{Sfig_two_vertex_state_circuit}, where we slightly
detune the group of qubits with a shorter duration to stop its entanglement evolution while waiting for the other group to finish the evolution under $H_{5q}$.

\subsection{1.4. Quantum state tomography}
\begin{figure*}[!htbp]
	\centering
	\includegraphics[width=\textwidth]{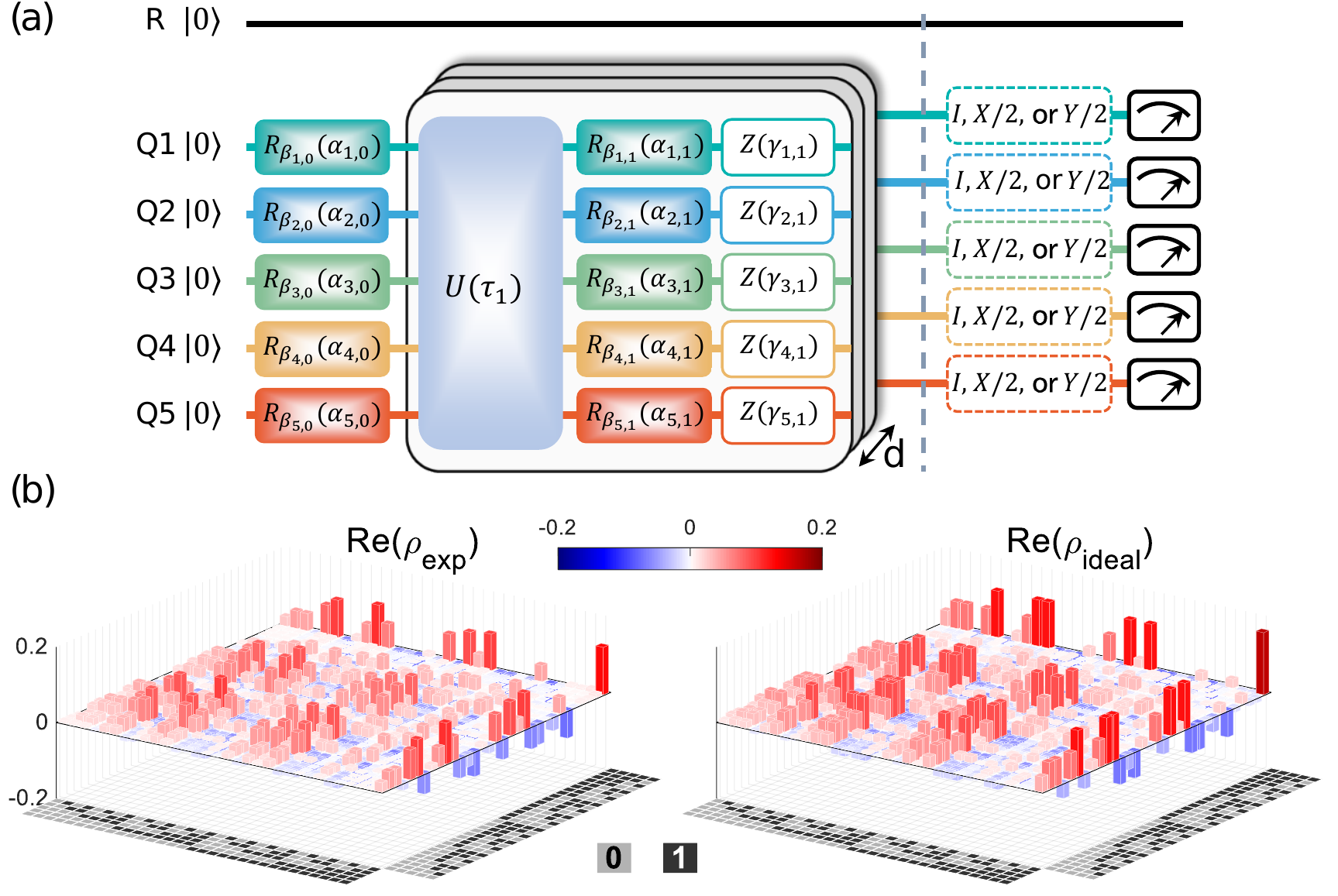}
	\caption{\label{Sfig_single_vertex_state_tomo}\footnotesize{
			(a) Sequence diagrams to generate the 5-qubit vertex state followed by characterizations via QST.
			(b) Real parts of the experimental (left) and ideal (right) density matrices of the 5-qubit vertex state for comparison. 
			All imaginary elements of the experimental density matrix are measured to be less than 0.03. 
	}}
\end{figure*}

We use quantum state tomography (QST) to characterize the experimentally generated 5-qubit vertex state. 
As illustrated in the sequence diagram in Fig.~\ref{Sfig_single_vertex_state_tomo}(a), a Pauli gate set selected from $\{I, X/2, Y/2\}^{\otimes 5}$
is inserted between the state preparation and the multiqubit readout, resulting in $2^5$ probabilities $\{P_{00000}, P_{00001}, \cdots, P_{11111}\}$.
With $3^5$ choices of the Pauli gate sets, we obtain a total of $3^5 \times 2^5$ measured probabilities, based on which we 
utilize convex optimization to extract the density matrix which is constrained to be Hermitian, normalized, and positive semidefinite. 
The experimentally obtained density matrix $\rho_{\mathrm{exp}}$ of the 5-qubit vertex state is depicted in Fig.~\ref{Sfig_single_vertex_state_tomo}(b), with 
the state fidelity of $\mathrm{Tr}\left(\rho_{\mathrm{ideal}} \cdot \rho_{\mathrm{exp}}\right) = 0.832\pm0.005$, where $\rho_{\mathrm{ideal}}$ is from Tab.~\ref{Stab_vertex_state_coefficients_1}. 

\subsection{1.5 Two-qubit controlled phase gate $C_\varphi$ and quantum process tomography}
\begin{figure}
	\centering
	\includegraphics[width=3.4in]{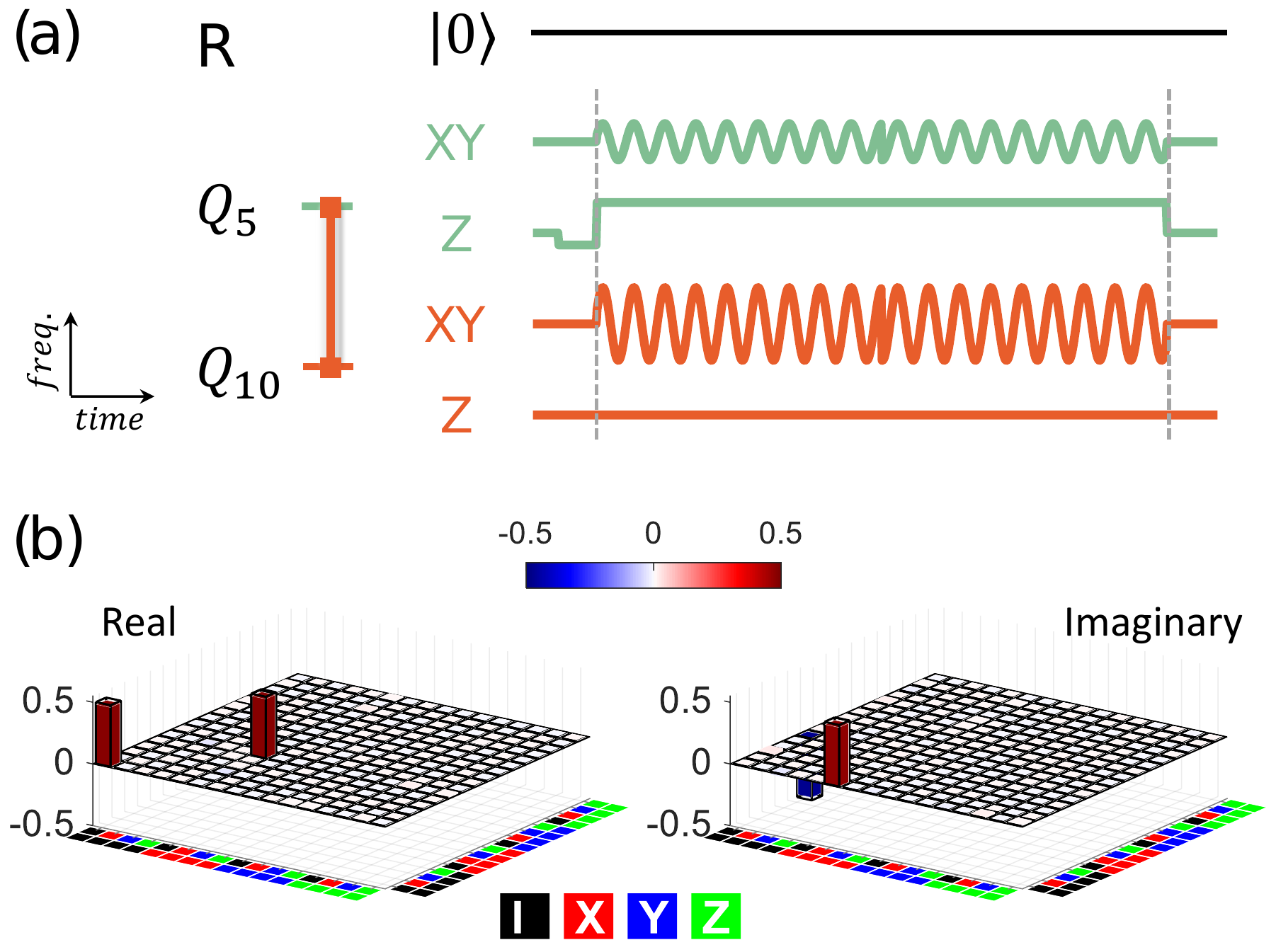}
	\caption{\label{Sfig_two_qubit_gate} \footnotesize{
			(a) Pulse sequence to generate the controlled phase gate $C_\varphi$ acting on the two qubits ($Q_5$ and $Q_{10}$), which interact with each other via mediation by resonator $R$.
			(b) Real and imaginary parts of the experimental process matrix characterizing $C_\varphi$ (solid bars) in comparison with the ideal ones (black frames). The color code for the Pauli basis $\{I,X,Y,Z\}$ is
			shown at the bottom.
	}}
\end{figure}

As mentioned in the main text, gluing two vertex states requires a two-qubit controlled phase gate ($C_\varphi$) acting on $Q_{5}$ and $Q_{10}$. 
Here we follow the protocol outlined in Ref.~\cite{PhysRevLett.121.130501} to implement $C_\varphi$, with the pulse sequence illustrated in Fig.~\ref{Sfig_two_qubit_gate}(a). 
While $Q_5$ and $Q_{10}$ are originally at their idle frequencies $5.080$~GHz and $5.043$~GHz, respectively, a 15~ns-long square pulse ($Z$ phase gate) is first applied to $Q_5$ to align its $x$-axis of the Bloch sphere to $Q_{10}$'s referenced to the rotating frame at $5.043$~GHz. 
Then $Q_{5}$ is biased to $5.043$~GHz for an on-resonant interaction with $Q_{10}$, while $XY$ microwaves with the driving amplitudes of approximately $2.96$ and $12.74$~MHz are applied to $Q_5$ and $Q_{10}$, respectively. Both microwave phases are initialized to be aligned to $x$-axis of the Bloch sphere and then inverted at the middle point of the interaction duration. This on-resonant interaction with a duration around $228$~ns fulfills a $C_\varphi$ in the dressed state basis.

To characterize $C_\varphi$, we perform quantum process tomography by preparing 36 distinct input states generated 
by operations from the gate set $\{I, \pm X/2, \pm Y/2, X\}^{\otimes 2}$, and measure the resulting output states by QST. 
Based on the QST density matrices of the input and output states, we are able to estimate the experimental matrix $\chi_{\mathrm{exp}}$ describing the $C_\varphi$ process, yielding the process fidelity of $\mathrm{Tr}\left(\chi_{\mathrm{ideal}}\chi_{\mathrm{exp}}\right) = 0.958 \pm 0.003$, where $\chi_{\mathrm{ideal}}$ refers to the ideal matrix. 

\section{2. Details of Quantum tetrahedra}

\subsection{2.1. Quantum tetrahedra and the spin-network states}
Denote $\mathcal{H}_{j}$ as a $2j+1$-dimensional Hilbert space consisting all the spin $j$ states. A quantum tetrahedron \cite{baez1999quantum,Rovelli_2006,LS,Rovelli:2015fwa} is a tensor state $\ket{e}$ in $\otimes_{i=1}^4 \mathcal{H}_{j_i}$ satisfying 
\be\label{eq:closure}
(\hat{\mathbf{J}}_1+\hat{\mathbf{J}}_2+\hat{\mathbf{J}}_3+\hat{\mathbf{J}}_4)\ket{e}=0.
\ee
Here the vector operators $\hat{\mathbf{J}}_i=(\hat{\mathbf{J}}_i^x,\hat{\mathbf{J}}_i^y,\hat{\mathbf{J}}_i^z)$ are the spin operators defined on spaces $\mathcal{H}_{j_i}$. These vector operators follows the relation $\hat{\mathbf{J}}_i\times\hat{\mathbf{J}}_i=\ii\hat{\mathbf{J}}_i$,where $\times$ stands for the vector product, and $[\hat{\mathbf{J}}_i,\hat{\mathbf{J}}_j]=0$ if $i\neq j$.

The equation \eqref{eq:closure} brings in the geometric interpretation of the quantum tetrahedron state $\ket{e}$ by the following points
\begin{enumerate}
	\item 
	Define a face normal operator as $\hat{\mathbf{E}}_{i}=8\pi \gamma l_p^2\hat{\mathbf{J}}_i$, where the Immirzi parameter $\gamma$ and the Planck length $l_p$ bring in the physical size of the face area \cite{Immirzi_1997}. The expectation values $\mathbf{E}_{i}=\bra {e} \hat{\mathbf{E}}_{i}\ket{e},\ (i=1\cdots4)$ can be interpreted as four face normals of a closed tetrahedron satisfying $\sum_{i=1}^4 \mathbf{E}_i=0$. 
	
	\item The expectation values of geometry operators, e.g.,the `cosine' dihedral angle operators $\widehat{\cos\theta}_{km}=\frac{4}{3}\ {\hat{\mathbf{J}}_{j_k}\cdot\hat{\mathbf{J}}^{j_m}}$, and the volume operator $\hat{V}=\frac{\sqrt{2}}{3} (\sqrt{8\pi\gamma }l_p)^3 \sqrt{|(\hat{\mathbf{J}}_1\times\hat{\mathbf{J}}_2)\cdot \hat{\mathbf{J}}_3|}$, provide the data that can reconstruct the `shape' of $e$. 
	\item The quantum fluctuations of the geometric quantities  make the `shape' of the quantum tetrahedron fuzzy. In the `large-spin' limit $j\to\infty$, these quantum fluctuations are negligible.
\end{enumerate}  

A quantum tetrahedron state is also called a rank-$4$ $\Su$-intertwiner. In general, a state $\ket{f}$ in $\otimes_{i=1}^n \mathcal{H}_{j_i}$, satisfying 
\[
\sum_{i=1}^n \hat{\mathbf{J}}_i\ket{f}=0,
\]
is called a rank-$n$ $\Su$-intertwiner. Geometrically, the state $\ket{f}$ can be considered as a quantum polyhedron.

In analog to the fact that any $3$-dimensional space can be triangulated and approximated by a large number of tetrahedra (polyhedral), the quantum states comprising many quantum tetrahedra (polyhedral) can be considered as quantized 3-dimensional spaces. Such states are called the spin-network states. A spin-networks state \cite{Rovelli_1995,Baez_1996,Ashtekar_1997,Haggard:2011qvx,Rovelli:2015fwa} can be represented as a graph made of many vertices and many links connecting the vertices. Each $4$-valent ($n$-valent) vertex is assigned with a rank-$4$ (rank-$n$) $\Su$-intertwiner as a quantum tetrahedron (polyhedron). Each link is labeled with a spin variable $j$ indicating the common boundary face of two quantum tetrahedra (polyhedral). In SFM, the spin-network states form a basis of the model's Hilbert space. The evolution of a spin-network state forms a quantum spacetime whose dynamical properties are described by the corresponding spin foam amplitude.

\subsection{2.2. Spin-$\frac{1}{2}$ quantum tetrahedra}
A quantum tetrahedron state $\ket{e}$ can be constructed by coupling four spin states denoted as $\ket{j_1,m_1}, \ket{j_2,m_2},\ket{j_3,m_3}$, and $\ket{j_4,m_4}$. Such coupling can be done in two steps. In the first step,  $\ket{j_1,m_1}, \ket{j_2,m_2}$ are coupled to be an intermediate state with spin $J_{12}$, while, $\ket{j_3,m_3}, \ket{j_4,m_4}$ are coupled to be a spin-$J_{34}$ state.  In the second step, the $J_{12}$ intermediate state and $J_{34}$ intermediate state are coupled into a final spin-$J$ state. 
\begin{displaymath}
\xymatrix{
	j_1\ar[dr]     &          & j_2\ar[dl] & & j_3\ar[dr] &          & j_4\ar[dl]    \\
	& J_{1\,2}\ar[drr] &            & &            & J_{3\,4}\ar[dll] &          \\
	&          &            &J &            &          &
}
\end{displaymath} 
The constraint \eqref{eq:closure} fixes the final spin $J$ to be $0$. Hence, $J_{12}$ must equal to $J_{34}$. 

In our experiment, all the $j_i$ equal to $\frac{1}{2}$. Then, by the trangle condition of coupling spins, the intermediate spin $J_{12}$ can be either $0$ or $1$. Thus the space of the spin-$\frac{1}{2}$ quantum tetrahedra is a $2$-dimensional space. 

Denoting $\ket{\uparrow\,}$ and $\ket{\downarrow\,}$ as the state  $\ket{\frac{1}{2},\frac{1}{2}}$ and $\ket{\frac{1}{2},-\frac{1}{2}}$, one basis of the spin-$\frac{1}{2}$ quantum tetrahedra space can be written as
\begin{equation}
\begin{split}
\ket{0}&=\frac{1}{2}\bigg(\ket{\uparrow\downarrow\,}-\ket{\downarrow\uparrow\,}\bigg)\bigg(\ket{\uparrow\downarrow\,}-\ket{\downarrow\uparrow\,}\bigg),\\
\ket{1}&=\frac{1}{\sqrt{3}}[\ket{\downarrow\downarrow\uparrow\uparrow\,}+\ket{\uparrow\uparrow\downarrow\downarrow\,}\\ 
&-\frac{1}{2}\big(\ket{\uparrow\downarrow\,}+\ket{\downarrow\uparrow\,}\big)\big(\ket{\uparrow\downarrow\,}+\ket{\downarrow\uparrow\,}\big)].
\end{split}
\end{equation}
Thus, a spin-$\frac{1}{2}$ quantum tetrahedron state $\ket{e}$ can be expressed as
\[
\ket{e}=\cos \frac{\theta}{2} \ket{0}+e^{\ii\phi}\sin \frac{\theta}{2}\ket{1},
\]
where $\theta$ and $\phi$ are the coordinates on the Bloch-sphere.

\subsection{2.3. Regular quantum tetrahedra}
In our paper, we use regular quantum tetrahedra as parts of the boundary state. A regular quantum tetrahedron is a quantum state that all of its spin variables are the same and all of its expectation values of the `cosine' dihedral angle operators are the same. In the `large-spin' limit, this state reconstructs a classical regular tetrahedron.

For a spin-$\frac{1}{2}$ quantum tetrahedron state $\ket{e}$, the expectation values of the `cosine' dihedral angle operators are given by 
\begin{align*}
\langle\widehat{\left.\cos \theta_{12}\right\rangle}&=\langle\widehat{\left.\cos \theta_{34}\right\rangle}=\cos ^{2} \frac{\theta}{2}-\frac{1}{3} \sin ^{2} \frac{\theta}{2},\\
\langle\widehat{\left.\cos \theta_{13}\right\rangle}&=\langle\widehat{\left.\cos \theta_{24}\right\rangle}=\frac{2}{3} \sin ^{2} \frac{\theta}{2}+\frac{2 \sqrt{3}}{3} \cos \frac{\theta}{2} \sin \frac{\theta}{2} \sin {\phi},\\
\langle\widehat{\left.\cos \theta_{14}\right\rangle}&=1-\langle\widehat{\left.\cos \theta_{23}\right\rangle}-\langle\widehat{\left.\cos \theta_{12}\right\rangle},
\end{align*} 
where $\theta_{ij}$ are the dihedral angles between face $i$ and face $j$, and $(\theta,\phi)$ is the Bloch-sphere coordinate of the $\ket{e}$ \cite{Li_2019}.

When $(\theta,\phi)=(\frac{\pi}{2},\frac{\pi}{2})$ and  $(\theta,\phi)=(\frac{\pi}{2},\frac{3\pi}{2})$, all the six dihedral angles are equal to each other. Hence $\ket{e^+}= \ket{0}+\ii \ket{1}$ and $\ket{e^-}= \ket{0}-\ii \ket{1}$ are two spin-$\frac{1}{2}$ regular quantum tetrahedra.

\subsection{2.4. Orientation of a quantum tetrahedron}
The states $\ket{e^\pm}$ are eigenstates of the volume operator
\begin{align*}
\hat{V}\ket{e^+}&=(\sqrt{8\pi\gamma} l_p)^3\frac{\sqrt{2}}{3}\sqrt{\frac{\sqrt{3}}{4}}\ket{e^+},\\
\hat{V}\ket{e^-}&=-(\sqrt{8\pi\gamma} l_p)^3\frac{\sqrt{2}}{3}\sqrt{\frac{\sqrt{3}}{4}}\ket{e^-}.
\end{align*}
In SFM, the plus and minus sign in the eigenvalues indicate the orientations of quantum tetrahedron \cite{rovelli2014covariant}. For a general quantum tetrahedron, the orientation depends on the sign of expectation value of volume square operator $\hat{V}$.  

In SFM, a boundary state made of quantum tetrahedra is called physical, if those quantum tetrahedra corresponds to the boundary of a $4$-dimensional simplical complex and the orientations of those quantum tetrahedra are consistent. For a spacetime atom, the boundary is orientation consistent if all of its boundary tetrahedra have the same orientation. For two neighboring spacetime atoms, we say that the boundary is orientation consistent if the orientations of the
tetrahedron shared by them are opposite respecting to the two atoms \cite{HZ,HZ1}.

Our experiment shows that the physical boundaries that are orientation consistently provide  much larger spin foam amplitudes than those provided by the orientation inconsistent boundaries. 

\section {3. The Ooguri model Spin Foam Model}

In Ooguri model \cite{doi:10.1142/S0217732392004171,Rovelli:2015fwa,Perez2012}, the vertex amplitude is given by
\begin{equation}\label{eq:OoguriAmp}
\begin{split}
A_v=&\prod_{f}(2j_f+1)\int_{\Su}\prod_{e} dg_{ve} \times\\
&\prod_{f}\bra{j_f,\vec{n}_{ef}}g_{ve}^{-1} g_{v e^\prime}  \ket{j_f,\vec{n}_{e^\prime f}}.
\end{split}
\end{equation}
where $dg_{e}$ are the $\Su$ Haar measures, $g_e$ are $\Su$ group elements assigned to boundary tetrahedra labeled by $e$, and $\ket{j_f,\vec{n}_{e f}}$ are the coherent states of the faces of the boundary tetrahedra. In our paper, the spin variables $j_f$ are $\frac{1}{2}$. Thus the vertex amplitude can be simplified as
\begin{equation}\label{eq:OoguriAmpS}
\begin{split}
A_v=&2^{10}\int_{\Su}\prod_{e} dg_{ve} \prod_{f}\bra{\vec{n}_{ef}}g_{ve}^{-1} g_{v e^\prime}  \ket{\vec{n}_{e^\prime f}},
\end{split}
\end{equation}
where the $\ket{\vec{n}_{ef}}$ stand for the coherent state $\ket{\frac{1}{2},\vec{n}_{e f}}$.
Using a graphic representation, the vertex amplitude is expressed as  \eqref{eq:Amp1}.
\begin{equation}\label{eq:Amp1}
\centering \includegraphics[width=3in]{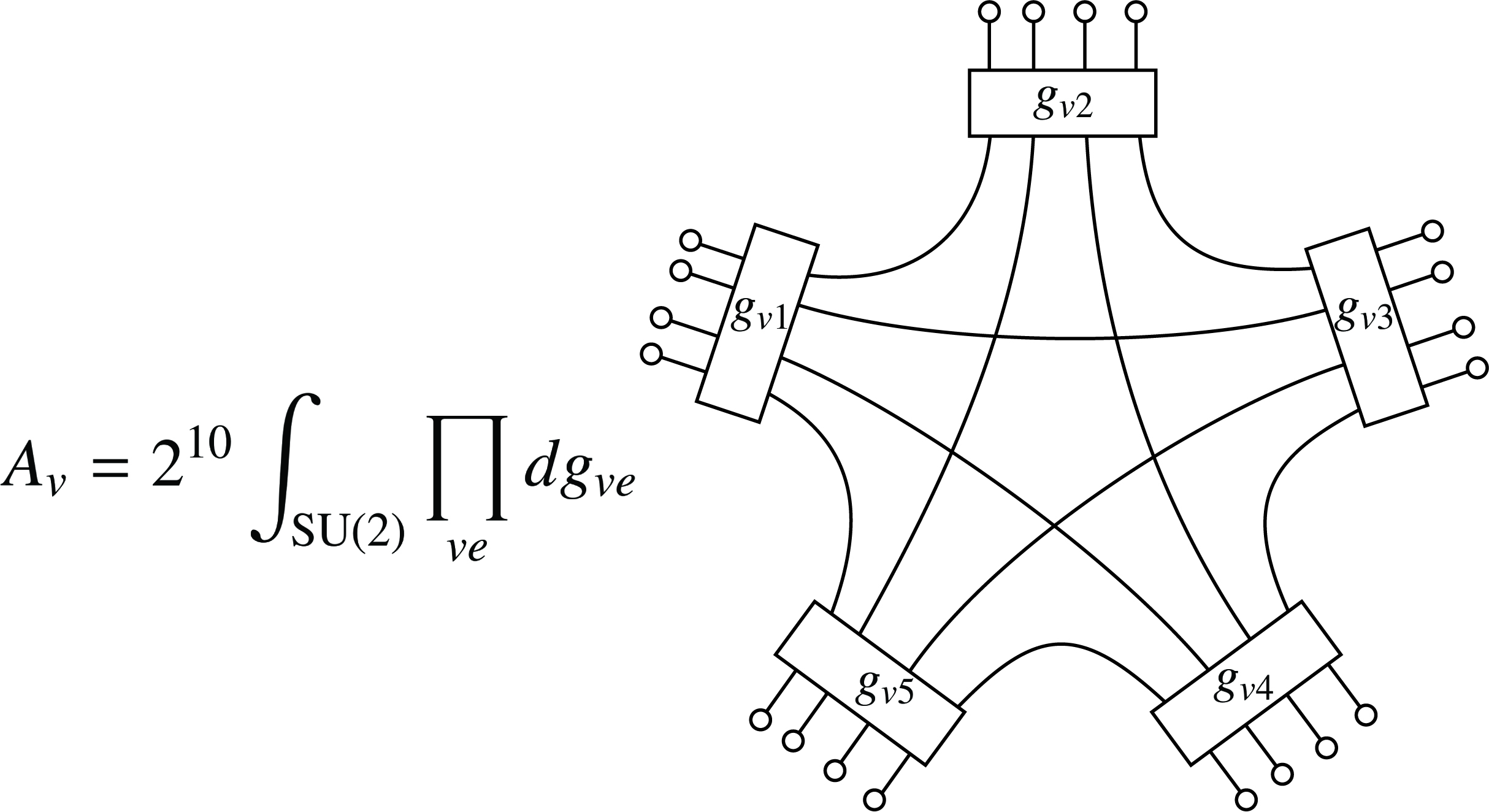}.
\end{equation}
Each circle in \eqref{eq:Amp1} stands for a state $\ket{\vec{n}_{ef}}$, each box indicates an group element $g_{ve}$, and each curve stands for an inner-product $\bra{\vec{n}_{ef}}g_{ve}^{-1} g_{v e^\prime} \ket{\vec{n}_{e^\prime f}}$. 
We perform the integrals over $\Su$ and get
\begin{equation}\label{eq:15jin}
\centering\includegraphics[width=2.5in]{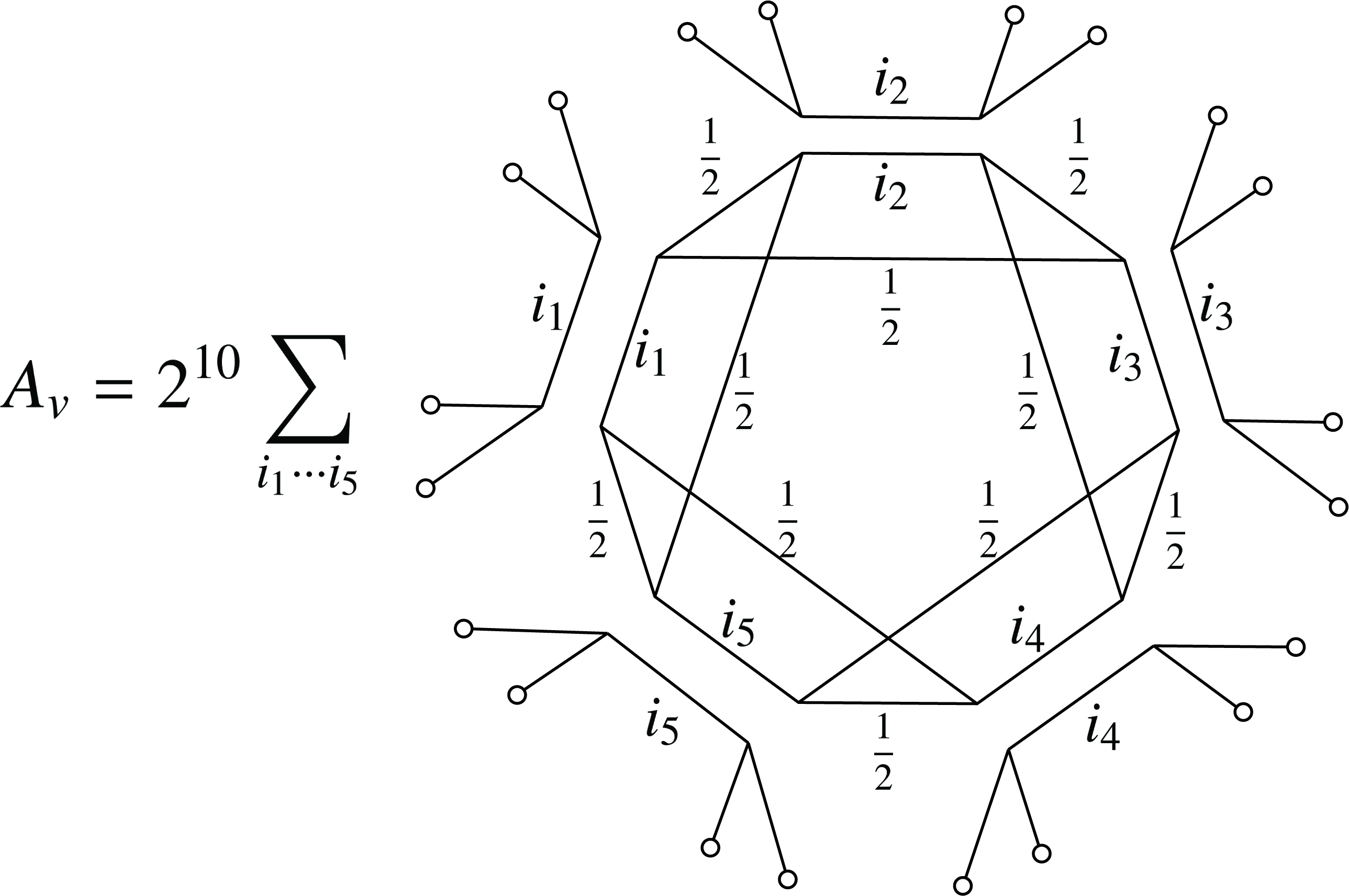}.
\end{equation}
In \eqref{eq:15jin}, the graphic notation 
\[
\centering\includegraphics[width=1.5in]{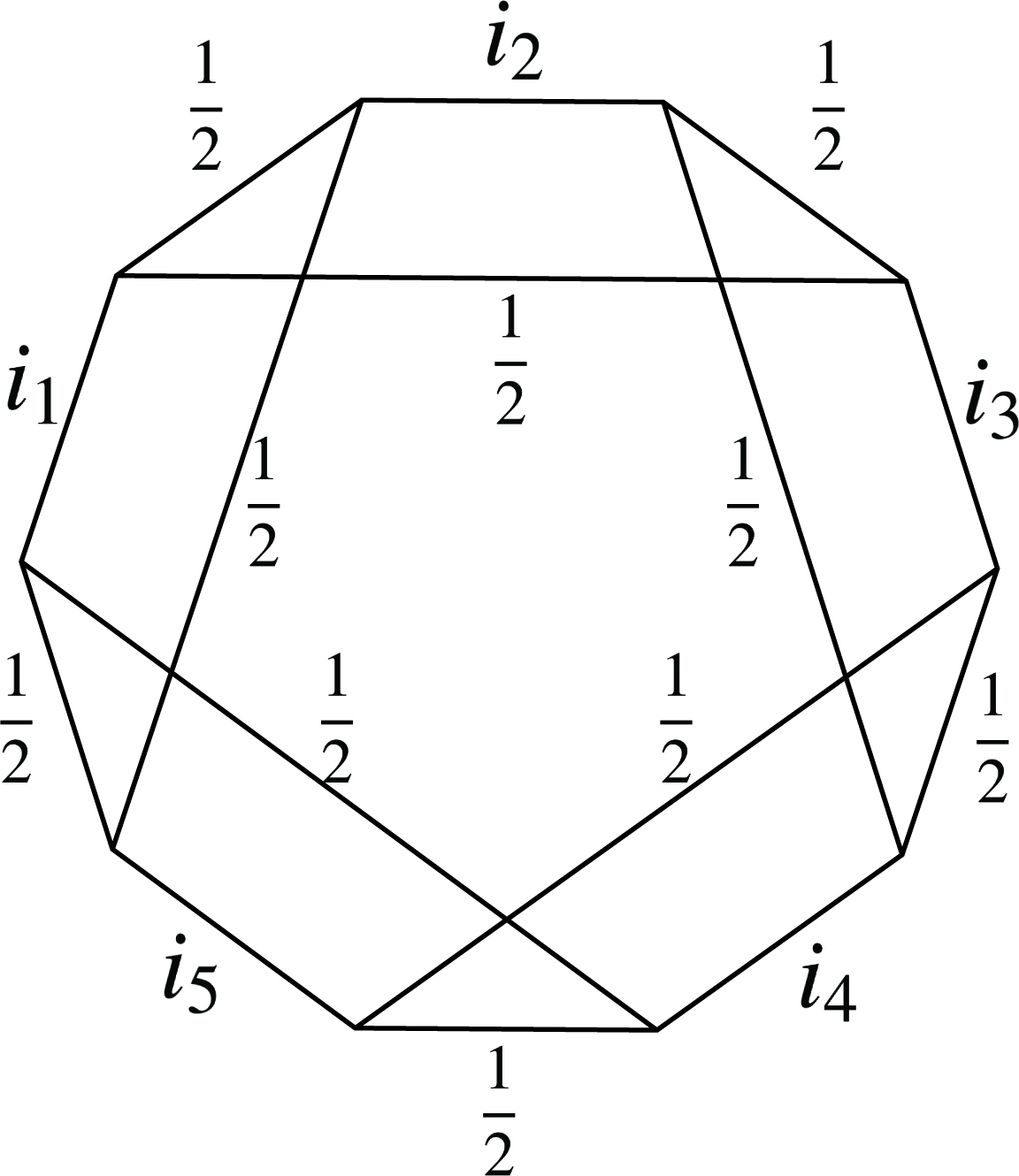}.
\]
is a $\{15j\}$-symbol which can be considered as a function of $15$ variables $f(j_1,\cdots,j_{10},i_1,\cdots,i_5)$. Ten of the variables are the spin variables (denoted as $j_k$) of the quantum tetrahedra and five of the variables are the intermediate spins (denoted as $i_e$) of the quantum tetrahedra. In our experiments, all $j_k$ equal to $\frac{1}{2}$. The intermediate spins $i_e$, as we mentioned in previous section, can be either $0$ or $1$. Thus, one can encode the information of this $\{15j\}$-symbol into a 5-qubit state $\ket{W}$, such that
\[
\ket{W}=\frac{1}{\mathcal{N}}\sum_{i_1,\cdots,i_5=00\cdots0}^{11\cdots1} f\left(\frac{1}{2},\cdots,\frac{1}{2},i_1,\cdots,i_5\right)\ket{i_1,\cdots,i_5},
\]
where $\mathcal{N}$ is the normalization factor. In our paper, we call $\ket{W}$ the vertex state. The components of $\ket{W}$ are given in TABLE \ref{Stab_vertex_state_coefficients_1}.


Denote a tensor state $\otimes_{i=1}^4|\vec{n}_{ei}\rangle$ as $\ket{e'}$.
Each 4-valent diagram 
\[
\centering\includegraphics[width=1.5in]{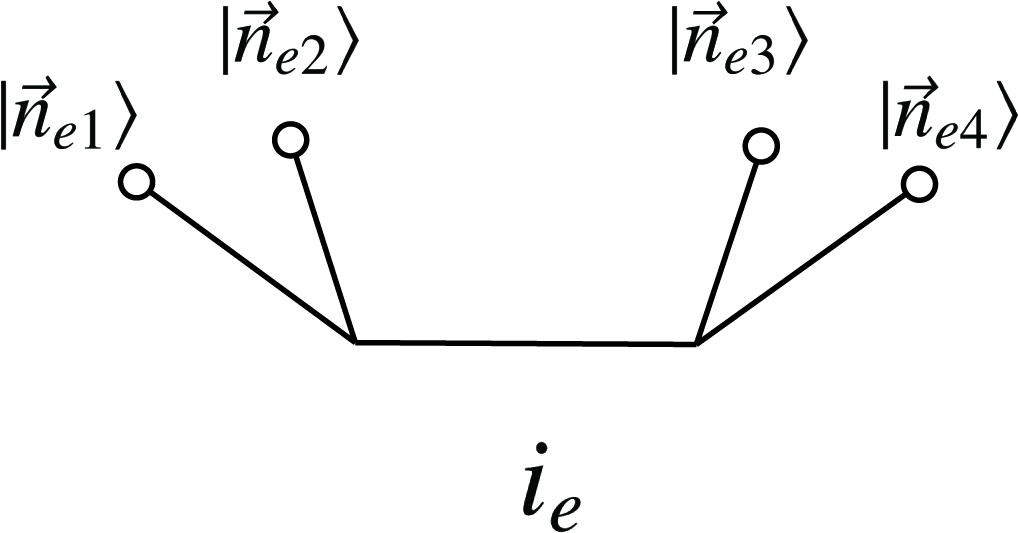}.
\]
in \eqref{eq:15jin} is a function $g(\vec{n}_{e1},\cdots,\vec{n}_{e4},i_e)$ whose value is given by
\[
g(\vec{n}_{e1},\cdots,\vec{n}_{e4},i_e)=\bra{i_e}e'\rangle,\, (i_e=0,1),
\]
where the states $\ket{0}$ and $\ket{1}$ are given in the previous section. Then, this 4-valent diagram can be considered as a projector that projects the $\ket{e'}$ to a quantum tetrahedra state
\[
\ket{e}=g(\vec{n}_{e1},\cdots,\vec{n}_{e4},0)\ket{0}+g(\vec{n}_{e1},\cdots,\vec{n}_{e4},1)\ket{1}.
\] 
satisfying \eqref{eq:closure}. 
So the vertex amplitude in \eqref{eq:15jin} can be written as an inner-product
\[
A_v=2^{10} \mathcal{Z} \bra{W}\otimes_{e=1}^5\ket{e},
\] 
where $\mathcal{Z}=0.62361$.

Multiple spacetime atoms are connected by identifying their common boundary. Graphically a two-vertex amplitude is shown as \eqref{eq:amp2}.
\begin{equation}\label{eq:amp2}
\centering\includegraphics[width=3.2in]{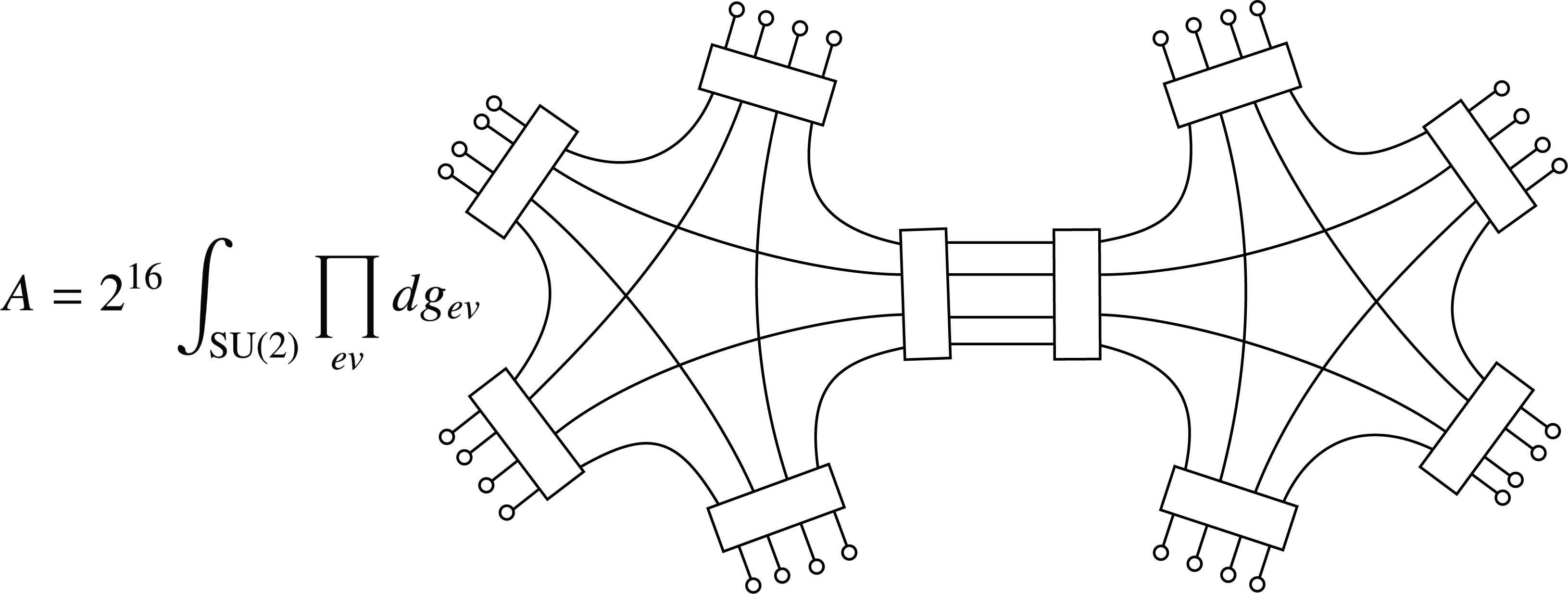}
\end{equation}
By doing the integrals, the amplitude becomes
\begin{equation}\label{eq:15jin2}
\centering\includegraphics[width=3in]{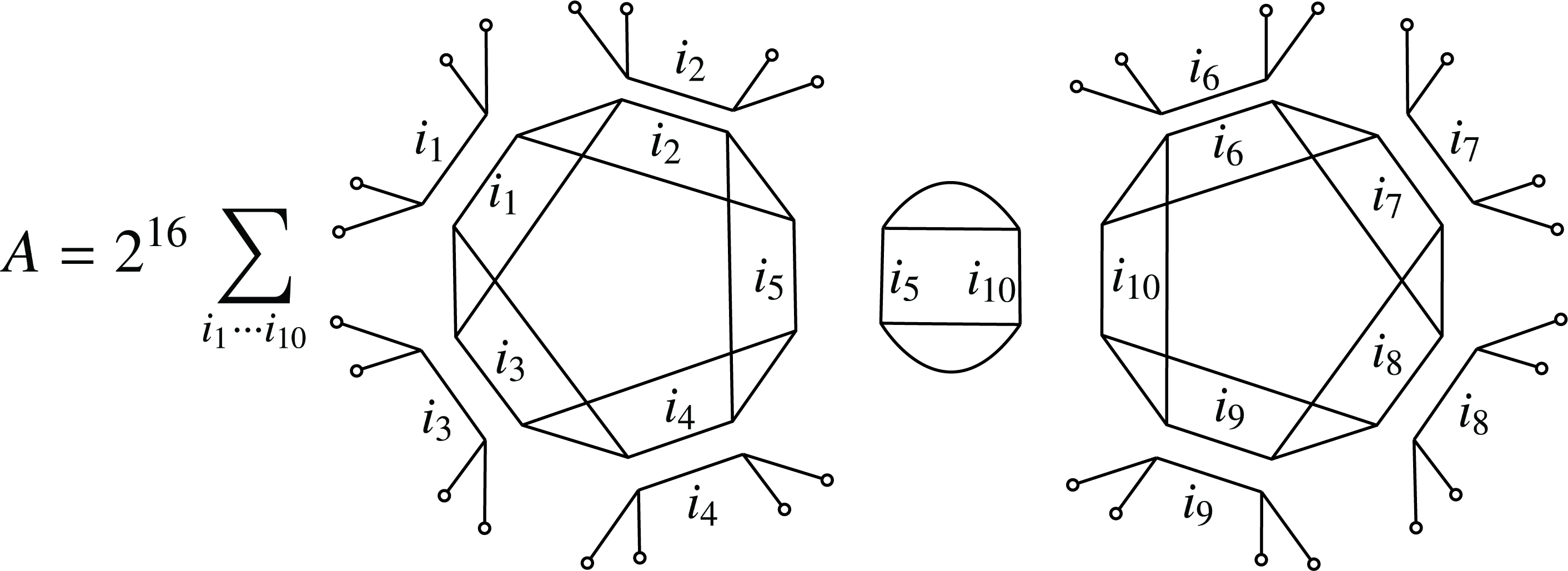},
\end{equation}
where the symbol
\[
\centering\includegraphics[width=0.4in]{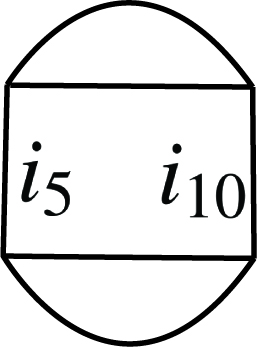}.
\] 
is $\delta_{i_5 i_{10}}$ when all spin variables $j$ are $\frac{1}{2}$. Thus the two-vertex state $\ket{W_d}$ is given by the diagram
\[
\centering\includegraphics[width=2.5in]{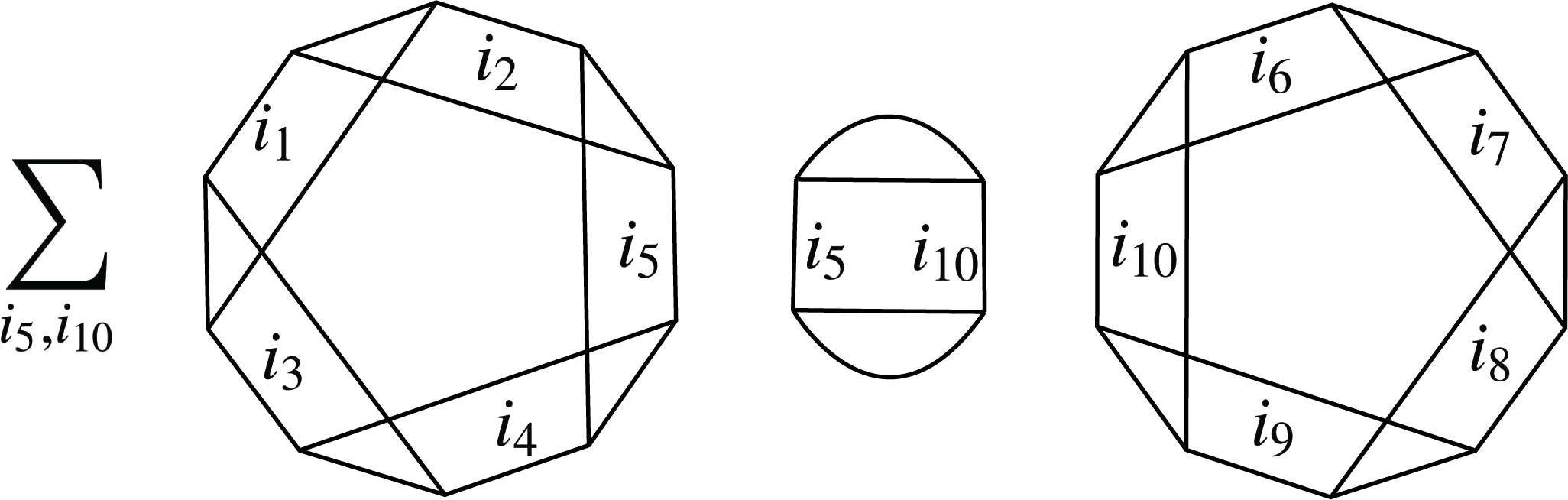},
\]
indicating that $\ket{W_d}$ is made by entangling two $\ket{W}$s.

Similar to the single atom case, the spin foam amplitude \eqref{eq:amp2} can be expressed as an inner-product
\[
A=\sqrt{2} (2)^{16} \mathcal{Z}^2  \bra{W_d}\otimes_{e=1}^8\ket{e},
\]
where the factor $\sqrt{2}$ comes from the normalization factor of the EPR state $\frac{\ket{00}+\ket{11}}{\sqrt{2}}$.

\section{4. The inner products}

In spin-$\frac{1}{2}$ cases, each $\ket{{e}_i}$ is a qubit state $\cos \frac{\theta_i}{2} \ket{0}+e^{\ii\phi_i}\sin \frac{\theta_i}{2}\ket{1}$ which can be generated by acting a rotation gate $U(\theta_i,\phi_i)$ on $\ket{0}$. Thus an inner-product between $\ket{{e}_i}$ and a qubit state $\ket{\psi}$ can experimentally measured by the following way. 
\begin{enumerate}
	\item Generate $\ket{\psi}$ in the system.
	\item Act $U^{-1}(\theta_i,\phi_i)$ on $\ket{\psi}$.
	\item Measure the probability of getting $\ket{0}$ as the output of the quantum gate. The square root of this probability provides the value of $\langle \psi | {e}_i \rangle$ up to a phase factor.
\end{enumerate}

In our experiments, vertex state $\ket{W}$ (two-vertex state $\ket{W_d}$) is a 
$5$ ($8$) qubits state. We act $5$ ($8$) inverse gates $U^{-1}(\theta_i,\phi_i)$ on $\ket{W}$ ($\ket{W_d}$) and measure the probability of getting all $\ket{0}$ as the output of the inverse gates as the modulus square of the spin foam amplitude $\langle{W}|{\Phi}\rangle$ ($\langle{W_d}|{\Phi}\rangle$)
\end{appendices}

\end{document}